\documentclass[11pt,a4paper,pdftex]{article}
\pdfoutput=1
\usepackage{jcappub}
\usepackage{multirow}
\usepackage{amsfonts}
\usepackage{ulem}
\usepackage{multirow}
\usepackage{caption}
\usepackage{subfig}
\usepackage{enumitem}
\usepackage{lipsum}
\usepackage{color}

\title{Testing the Kerr metric using X-ray reflection spectroscopy: spectral analysis of GX~339$-$4}

\author[b]{Jingyi~Wang,}
\author[a]{Askar~B.~Abdikamalov,}
\author[a]{Dimitry~Ayzenberg,}
\author[a,c,1]{Cosimo~Bambi,%
\note{Corresponding author}}
\author[d]{Thomas~Dauser,}
\author[e,d]{Javier~A.~Garc{\'\i}a,}
\author[c]{Sourabh~Nampalliwar,}
\author[f,b]{James~F.~Steiner}

\affiliation[a]{Center for Field Theory and Particle Physics, Department of Physics, Fudan University, 2005 Songhu Road, 200438 Shanghai, China}
\affiliation[b]{MIT Kavli Institute for Astrophysics and Space Research, MIT, 77 Massachusetts Avenue, Cambridge, MA 02139, USA}
\affiliation[c]{Theoretical Astrophysics, Eberhard-Karls Universit\"at T\"ubingen, Auf der Morgenstelle 10, 72076 T\"ubingen, Germany}
\affiliation[d]{Remeis Observatory \& ECAP, Universit\"{a}t Erlangen-N\"{u}rnberg, Sternwartstrasse 7, 96049 Bamberg, Germany}
\affiliation[e]{Cahill Center for Astronomy and Astrophysics, California Institute of Technology, 1216 East California Boulevard, Pasadena, CA 91125, USA}
\affiliation[f]{Harvard-Smithsonian Center for Astrophysics, 60 Garden Street, Cambridge, MA 02138, USA}

\emailAdd{bambi@fudan.edu.cn}

\abstract{Signatures of X-ray reprocessing (reflection) out of an accretion disk are commonly observed in the high-energy spectrum of accreting black holes, and can be used to probe the strong gravity region around these objects. In this paper, we extend previous work in the literature and we employ a full emission model for relativistic reflection in non-Kerr spacetime to demonstrate an approach that tests the Kerr black hole hypothesis. We analyze a composite spectrum obtained with the Proportional Counter Array in the Rossi X-ray Timing Explorer (RXTE), of the stellar-mass black hole GX~339$-$4 in its brightest hard state. With a remarkable sensitivity of $\sim0.1\%$ and 40~million counts in the $3-45$~keV band to capture the faint features in the reflection spectrum, we demonstrate that it is possible with existing data and an adequate model to place constraints on the black hole spin $a_*$ and the deformation parameter that quantifies the departure from the Kerr metric. Our measurement obtained with the best fit model, which should be regarded as principally a proof of concept, is $a_*=0.92^{+0.07}_{-0.12}$ and $\alpha_{13}=-0.76^{+0.78}_{-0.60}$ with a 90\% confidence level and is consistent with the hypothesis that the compact object in GX~339$-$4 is a Kerr black hole. We also discuss how the physical model choice and the emissivity profile adopted could make an impact on the constraints of $\alpha_{13}$ and spin. To enable Kerr metric test using X-ray reflection spectroscopy, it is essential to improve our astrophysical understanding of accreting black holes, e.g., the natures of accretion flow and corona.}

\keywords{gravity, astrophysical black holes, X-rays}

\begin{document}

\maketitle


\section{Introduction} \label{intro}

Einstein's theory of general relativity (GR hereafter) was proposed at the end of 1915~\cite{einstein} and is currently the standard framework for the description of gravitational fields and the chrono-geometrical structure of spacetime. While largely successful, application of Einstein's theory to our Universe has led to some unexplained phenomena, e.g., dark energy. Even on purely theoretical grounds, there are challenges like the presence of singularities and the difficulty in finding a good theory of quantum gravity. A number of extensions and modifications of GR have been proposed to deal with these challenges. Over the past century, there have been significant efforts to test the theory in weak gravitational fields, mainly with experiments in the Solar System and radio observations of binary pulsars~\cite{will}. GR has been highly successful in this regime and it is desirable that alternative theories match with GR in this regime. Thus, most of the alternative theories have the same predictions as GR for weak fields and present deviations only when gravity becomes strong. Thanks to the advent of new observational facilities, there is today an increasing interest in the possibility of testing GR in the strong gravity regime~\cite{ns, bh1,bh2,bh3,bh4,bh5}. Black holes are the most interesting candidates for such strong gravity tests.

Black holes are extremely compact objects in which matter has collapsed to a singularity. Today we have a body of observational evidence strongly supporting the existence of dark and compact objects within our Galaxy and across the Universe which can be interpreted as the black holes of Einstein's theory. They are found in various scenarios, peppered throughout each galaxy as remnants of stellar collapse~\cite{collapse},  within galactic centers~\cite{maoz}, and in mergers~\cite{ligo1,ligo2}. Around an astrophysical black hole, a flattened band of spinning matter (gas, dust, and other stellar debris) outside the event horizon called the accretion disk is formed. In 4-dimensional GR, the only stationary and asymptotically flat vacuum (electro-vacuum) black hole spacetime with a regular exterior region is described by the Kerr (Kerr-Newman) solution~\cite{kerr,knewman}. It is also intriguing that these black holes are extremely simple objects, being defined by not more than three parameters: mass $M$, spin angular momentum $J$, and electric charge $Q$~\cite{h0,h1,h2,h3}.

It is a remarkable fact that the spacetime metric around astrophysical black holes is well described by the Kerr solution. As soon as a black hole is formed, initial deviations from the Kerr metric are quickly radiated away with the emission of gravitational waves~\cite{price1,price2}. The equilibrium electric charge is extremely small for macroscopic objects and completely negligible for the metric of the spacetime~\cite{bambi2009black}. Accretion disks typically have a mass of many orders of magnitude smaller than the central object and their impact on the background metric can be safely ignored~\cite{barausse2014can,naokidisk}. Eventually, macroscopic deviations from the Kerr spacetime would only be possible in the presence of new physics, such as classical extensions of GR~\cite{extentions_GR}, macroscopic quantum gravity effects~\cite{dvali2013_quantum_effects2,dvali2013_quantum_effects1,giddings2014_quantum_effects}, or in the presence of exotic matter~\cite{herdeiro2015_exotic_field,herdeiro2016_exotic_field}. These deviations would affect the spacetime in a variety of ways and by probing the spacetime geometry, one can identify the deviations and thus test the extensions and modifications of GR.

Currently, there are two approaches for probing the spacetime geometry: electromagnetic radiation~\cite{em1,em2,em3} and gravitational waves~\cite{gw1,gw2}. With the electromagnetic approach, currently there are two leading techniques to probe the spacetime geometry around a black hole: the study of the thermal spectrum of thin disks (continuum-fitting method)~\cite{cfm1,cfm2,cfm3} and the analysis of the relativistically smeared reflection spectrum of thin disks (reflection method, often called the iron line method)~\cite{i1,i2,i3}. Both techniques have been developed for measuring black hole spins under the assumption of Kerr background and can be naturally extended for testing the Kerr metric~\cite{t1,t2,t3,t4,t5,t6,t7,t8}. The reflection method has some advantages over the study of the thermal spectrum. The former can be easily applied to both stellar-mass and supermassive black holes, while the latter is only observable in stellar-mass black hole systems. The simple shape of the thermal continuum is advantageous for measuring spin (when assuming a Kerr geometry and fitting two parameters), but not suitable for probing higher order structures including searches for deviations from Kerr. In contrast, the reflection spectrum has a distinct structure with a number of prominent features, e.g. fluorescent lines and absorption edges. Thus, in the presence of high quality data and the correct astrophysical model, the reflection method is potentially quite a powerful tool to constrain the metric around black holes (see, for instance, Ref.~\cite{jiachen}). The current limitation of the technique is mainly the uncertainty in the astrophysical model (e.g., the coronal/disk geometry and dynamics of the corona).

In Refs.~\cite{bambi2016,abdikamalov2019}, we have presented {\sc relxill\_nk}, which is an extension of the relativistic reflection model {\sc relxill}~\cite{garcia2011x,garcia2013x,garcia2014improved,dauser2010broad,dauser2013irradiation,dauser2014role} to non-Kerr spacetimes. In the past couple of years, we have applied {\sc relxill\_nk} to XMM-Newton, Suzaku, and NuSTAR observations of stellar-mass~\cite{xu2018study,zhang2019kerr,liu19nk,zhang19nk} and supermassive black holes~\cite{bh5,ashutosh,ashu1,ashu2} to test the Kerr hypothesis. In the case of supermassive black holes, our tests have always confirmed the Kerr nature of the accreting object, in some cases with quite stringent constraints~\cite{ashu1,ashu2}. While it is reasonable to expect that the spacetime metric around astrophysical black holes is at least close to the Kerr solution, as predicted by GR, such results were not obvious a priori because our model has a number of simplifications introducing systematic uncertainties. In particular, it is well know that some sources are characterized by a high mass accretion rate, which makes their accretion disk thick, while our reflection model employs the Novikov-Thorne set-up for thin disks. On the contrary, we have often found challenging to test the Kerr hypothesis with stellar-mass black holes; see, e.g., Refs.~\cite{zhang2019kerr,liu19nk,zhang19nk}.

In this paper, we test the hypothesis that the background metric around an astrophysical black hole is described by the Kerr solution by applying {\sc relxill\_nk} to one more stellar-mass black hole, GX~339$-$4, to further investigate whether it is indeed true that this class of objects is less suitable for testing the Kerr hypothesis. Our study on GX~339$-$4 confirms such a claim, because we get model-dependent constraints; that is, if we make different astrophysical assumptions, we may find different results (i.e., either strong or weak constraints on possible non-Kerr features, depending on the assumptions, and in some cases we do not recover the Kerr solution at a high confidence level), which is something that we would not like to have in a test of GR.

The content of the paper is as follows. In Section~\ref{kerr}, we review parametrically deformed Kerr metrics and describe the non-Kerr metric that we implemented in our reflection model. We review the technique of X-ray reflection spectroscopy and our reflection model in Section~\ref{reflection}. Section~\ref{obs} describes the observations, data reduction, and outlines our procedure for combining the individual spectra into the composite spectrum. Our fitting procedure and results of the constraint of the deformation parameter and the spin parameter appear in Section~\ref{fitting}. We draw our conclusions in Section~\ref{conclusion}.

Throughout the paper, we employ units in which $G_{\rm N} = c = 1$ and the convention of a metric with signature ($-+++$).


\section{Testing the Kerr Paradigm} \label{kerr}

There are two natural strategies to test the Kerr black hole hypothesis. In the top-down approach, we begin with a specific modified theory of gravity which has solutions mimicking black holes. These solutions' differences from the Kerr solution are parametrized by one or more deformation parameters. In this approach, testing astrophysical data for the presence of deformation parameters amounts to testing the specific modified theory. This approach is the most logical, but it typically presents two problems. Firstly, there are a large number of modified theories of gravity and there is no clear preference for any one of them, so the search for deformation parameters has to be repeated for every theory. Secondly, black hole solutions are known in very few modified theories of gravity, and in many of those only the non-rotating and the slow-rotating solutions, while fast rotating black hole solutions, preferred observationally, are known only in exceptional cases.

In the bottom-up approach, we consider a phenomenological parameterization of the Kerr metric, which in principle should be able to describe the spacetime of any possible black hole in any possible gravity theory (in practice, bottom-up metrics describe black hole metrics with a range that encompasses a variety of, but not all, alternative theories). A number of deformation parameters are used to quantify possible deviations from the Kerr metric, and analysis of astrophysical data constrains potential deviations from the Kerr solution in a model-independent way.

The bottom-up approach is analogous to the Parametrized Post-Newtonian (PPN) formalism~\cite{will} that has been successfully employed to test the Schwarzschild solution in the weak field limit with Solar System experiments over the past several decades. Within the PPN framework, we write the most general static, spherically symmetric, and asymptotically flat metric in terms of the expansion parameter $M/r$ , where $M$ is the mass of the central object and $r$ is some radial coordinate. When arranged in the Schwarzschild coordinates, the line element
reads
\begin{align}
ds^2=&-\left[1-\frac{2M}{r}+(\beta-\gamma)\frac{2M^2}{r^2}+\ldots\right]dt^2\nonumber\\
&+\left(1+\gamma\frac{2M}{r}+\ldots\right)dr^2+r^2d\theta^2 +r^2\sin^2\theta d\phi^2 \, .
\end{align}
$\beta$ and $\gamma$ are free parameters to be measured by experiments. The only spherically symmetric vacuum solution of Einstein's equations is the Schwarzschild metric and it requires $\beta=\gamma=1$. Other theories of gravity may have a different spherically symmetric vacuum solution, and in this case $\beta$ and $\gamma$ may not be exactly 1. Current observational data in the Solar System provide the following constraints on $\beta$ and $\gamma$~\cite{bertotti2003test,williams2004progress}
\begin{equation}
|\beta -1| < 2.3 \cdot10^{-4} \, , \qquad |\gamma - 1| < 2.3 \cdot 10^{-5} \, ,
\end{equation}
confirming the validity of the Schwarzschild solution in the weak field limit within the precision of current observations.

There are several proposals for bottom-up metrics, the most extensively employed ones are the Johannsen-Psaltis metric~\cite{JPmetric}, the Johannsen metric~\cite{johannsen_metric}, and the Konoplya-Rezzolla-Zhidenko metric~\cite{KRZmetric}. Here we adopt the Johannsen metric to describe the spacetime geometry. Its most important properties are that: $i)$ the metric is regular (no naked singularities or closed time-like curves) everywhere on and outside of the event horizon, just like the Kerr metric, and $ii)$ it was explicitly shown that it is able to recover some black hole solutions in alternative theories of gravity for suitable choices of the free parameters. In Boyer-Lindquist coordinates, the line element of the Johannsen metric reads~\cite{johannsen_metric}
\begin{align}
ds^2=&-\frac{\tilde{\Sigma}\left(\Delta-a^2A_2^2\sin^2\theta\right)}{B^2}dt^2
-\frac{2a\left[\left(r^2+a^2\right)A_1A_2-\Delta\right]\tilde{\Sigma}\sin^2\theta}{B^2}dtd\phi \nonumber\\
&+\frac{\tilde{\Sigma}}{\Delta A_5}dr^2+\tilde{\Sigma}d\theta^2
+\frac{\left[\left(r^2+a^2\right)^2A_1^2-a^2\Delta\sin^2\theta\right]\tilde{\Sigma}\sin^2\theta}{B^2}d\phi^2
\end{align}
where
\begin{align}
&a=J/M \, , \qquad B=\left(r^2+a^2\right)A_1-a^2A_2\sin^2\theta \, , \nonumber\\
& \tilde{\Sigma}=\Sigma+f \, , \qquad \Sigma=r^2+a^2\cos^2\theta \, , \nonumber\\ 
& \Delta=r^2-2Mr+a^2 \, .
\end{align}
and the four free functions $f$, $A_1$, $A_2$, and $A_5$ are
\begin{align}
& f = \sum^{\infty}_{n=3}\epsilon_n\frac{M^n}{r^{n-2}} \, , \nonumber\\
& A_1 = 1 + \sum_{n=3}^{\infty}\alpha_{1n}\left(\frac{M}{r}\right)^n \, , \nonumber\\
& A_2 = 1 + \sum_{n=2}^{\infty}\alpha_{2n}\left(\frac{M}{r}\right)^n \, , \nonumber\\
& A_5 = 1 + \sum_{n=2}^{\infty}\alpha_{5n}\left(\frac{M}{r}\right)^n \, .
\end{align}
The metric elements depend on the mass and spin of the black hole as well as on four free functions that measure potential deviations from the Kerr solution. The first order deformation parameters in these free functions are $\epsilon_3$, $\alpha_{13}$, $\alpha_{22}$, and $\alpha_{52}$. This metric exactly reduces to the Kerr solution for $\epsilon_3=\alpha_{13}=\alpha_{22}=\alpha_{52}=0$.

In the Kerr metric, the condition for the existence of an event horizon is $| a_* | \le 1$, where $a_* = a/M = J/M^2$ is the dimensionless spin parameter. If $| a_* | > 1$, there is no horizon and the singularity at the origin $r=0$ is naked. In the Johannsen spacetime, we still have the condition $| a_* | \le 1$. Moreover, in order to exclude a violation of Lorentzian signature or the existence of closed time-like curves in the exterior region, we have to impose that the metric determinant is always negative and that $g_{\phi\phi}$ is never negative for radii larger than the radius of the event horizon. These conditions lead to the following constraints on the first-order deformation parameters~\cite{johannsen_metric}
\begin{align}
\label{eq:boundary}
\alpha_{13}, \epsilon_{3} & \geq - \left( 1 + \sqrt{1 - a_*^2} \right)^3 \, , \\
\alpha_{22}, \alpha_{52} & \geq - \left(1 + \sqrt{1 - a_*^2} \right)^2 \, .
\end{align}

In the present work, as a first step we only explore a non-vanishing $\alpha_{13}$, and we set to zero all other deformation parameters. As shown in Ref.~\cite{bambi2016}, $\alpha_{13}$ has the strongest impact on the iron line profile and on the whole reflection spectrum among the deformation parameters, and it is thus the most suitable to illustrate the potential capabilities of X-ray reflection spectroscopy to constrain possible deformations from the Kerr geometry. However, every deformation parameter changes the metric in a different way, has its own impact on the reflection spectrum, and therefore a model-independent confirmation of the Kerr nature of black holes would require to constrain all possible deformation parameters.

It can be helpful to have at least a qualitative idea of the role of $\alpha_{13}$ on the spacetime geometry around black holes. Since $\alpha_{13}$ enters the metric coefficients $g_{tt}$, $g_{t\phi}$, and $g_{\phi\phi}$, it alters the structure of our infinitesimally-thin accretion disk through the gas angular velocity $\Omega$
\begin{equation}
\Omega = \frac{- \partial_r g_{t\phi} \pm \sqrt{\left( \partial_r g_{t\phi} \right)^2 
- \left( \partial_r g_{tt} \right)\left( \partial_r g_{\phi\phi} \right)}}{\partial_r g_{\phi\phi}} \, ,
\end{equation}
and the position of the innermost stable circular orbit (ISCO), which is determined by effective potential
\begin{equation}
V_{\rm eff} = \frac{E^2 g_{\phi\phi} + 2 E L_z g_{t\phi} + L_z^2 g_{tt}}{g_{t\phi}^2 - g_{tt} g_{\phi\phi}} - 1 \, ,
\end{equation}
where $E$ and $L_z$ are, respectively, the conserved energy and angular momentum per unit rest-mass. See Ref.~\cite{bambi2016} for the details and all the calculations. $\alpha_{13}$ also changes the photon redshift, which is given by
\begin{equation}
g = \frac{\sqrt{- g_{tt} - 2 \Omega g_{t\phi} - \Omega^2 g_{\phi\phi} }}{1 - \lambda \Omega} \, ,
\end{equation}
where $\lambda = -k_\phi/k_t$ and $k_t$ and $k_\phi$ are, respectively, the $t$- and $\phi$-component of the conjugate 4-momentum of the photon, which are constants of motion along the photon trajectory. Roughly speaking, $\alpha_{13} > 0$ ($< 0$) increases (decreases) the strength of the gravitational force with respect to the standard GR case ($\alpha_{13} = 0$). As we will show in the next section, $\alpha_{13} > 0$ ($< 0$) reduces (enhances) the extension of the low energy tail of the iron line: the main impact of a stronger (weaker) gravitational force is to move the ISCO radius to higher (lower) values, and thus to reduce (increase) the number of strongly redshifted photons.

Since the metric is singular for $B = 0$, in our study we impose that $B$ never vanishes for radii larger than the event horizon, and we obtain the following constraint
\begin{align}
\label{eq:boundary-2}
\alpha_{13} \geq - \frac{1}{2} \left( 1 + \sqrt{1 - a_*^2} \right)^4 \, ,
\end{align}
which is stronger than the requirement in Eq.~(\ref{eq:boundary}).


\section{X-ray Reflection Spectroscopy}\label{reflection}

\begin{figure}[t]
\begin{center}
\includegraphics[type=pdf,ext=.pdf,read=.pdf,width=6.5cm]{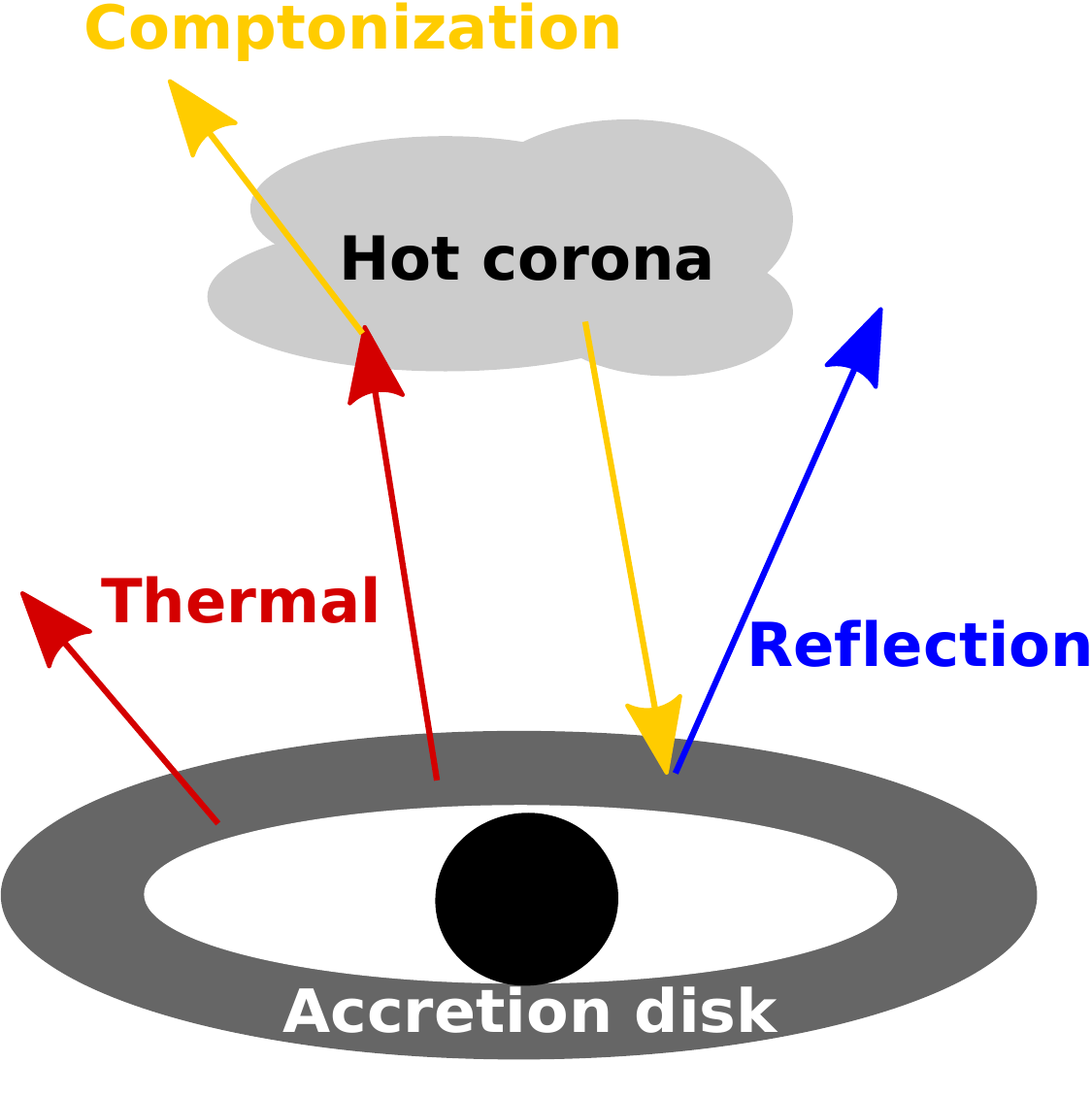}
\end{center}
\caption{A cartoon illustrating the process which gives rise to reflection emission from a disk-coronal geometry. The three main spectral components are indicated by different colored arrows. Thermal component: multi-temperature black body radiation from the disk. Comptonization component: inverse Compton scattering of the thermal photons off free electrons in a hot plasma cloud called the ``corona". Reflection component: the Comptonized photons illuminate the disk, and interactions including Compton scattering and photoelectric absorption generate the reflection emission. See the text for more details. \label{fig:corona-disk}}
\end{figure}

The standard picture for an accreting black hole system involves an accretion disk and a compact corona of hot electrons (see Fig.~\ref{fig:corona-disk} and, e.g., Ref.~\cite{annphi} for more details). The accretion disk emits as a blackbody locally, and as a multi-temperature blackbody when integrated radially. The temperature of the disk depends on the black hole mass, the mass accretion rate, the position in the disk, and at some level on the spin and possible deformation parameters. For a black hole accreting at $\sim 10$\% of its Eddington limit, the thermal emission of the inner part of the accretion disk is in the soft X-ray band ($\sim1$~keV) for stellar-mass black holes and in the optical/UV band ($1 \sim 10$~eV) for the supermassive ones. Inverse Compton scattering of the thermal photons from the accretion disk off free electrons in a hotter ($\sim100$ keV) plasma cloud, called the ``corona", generates a Comptonization component, which is often modeled with a cut-off power-law spectrum~\cite{Sunyaev:1979nz}. A fraction of the Comptonized photons illuminate the disk. The interaction of these photons with the material of the accretion disk, which mainly includes Compton scattering and photoelectric absorption followed by fluorescent line emission or Auger de-excitation, produces the reflection spectrum~\cite{George:1991jj,Ross:2005dm}.

\begin{figure*}
\begin{center}
\includegraphics[type=pdf,ext=.pdf,read=.pdf,width=7cm]{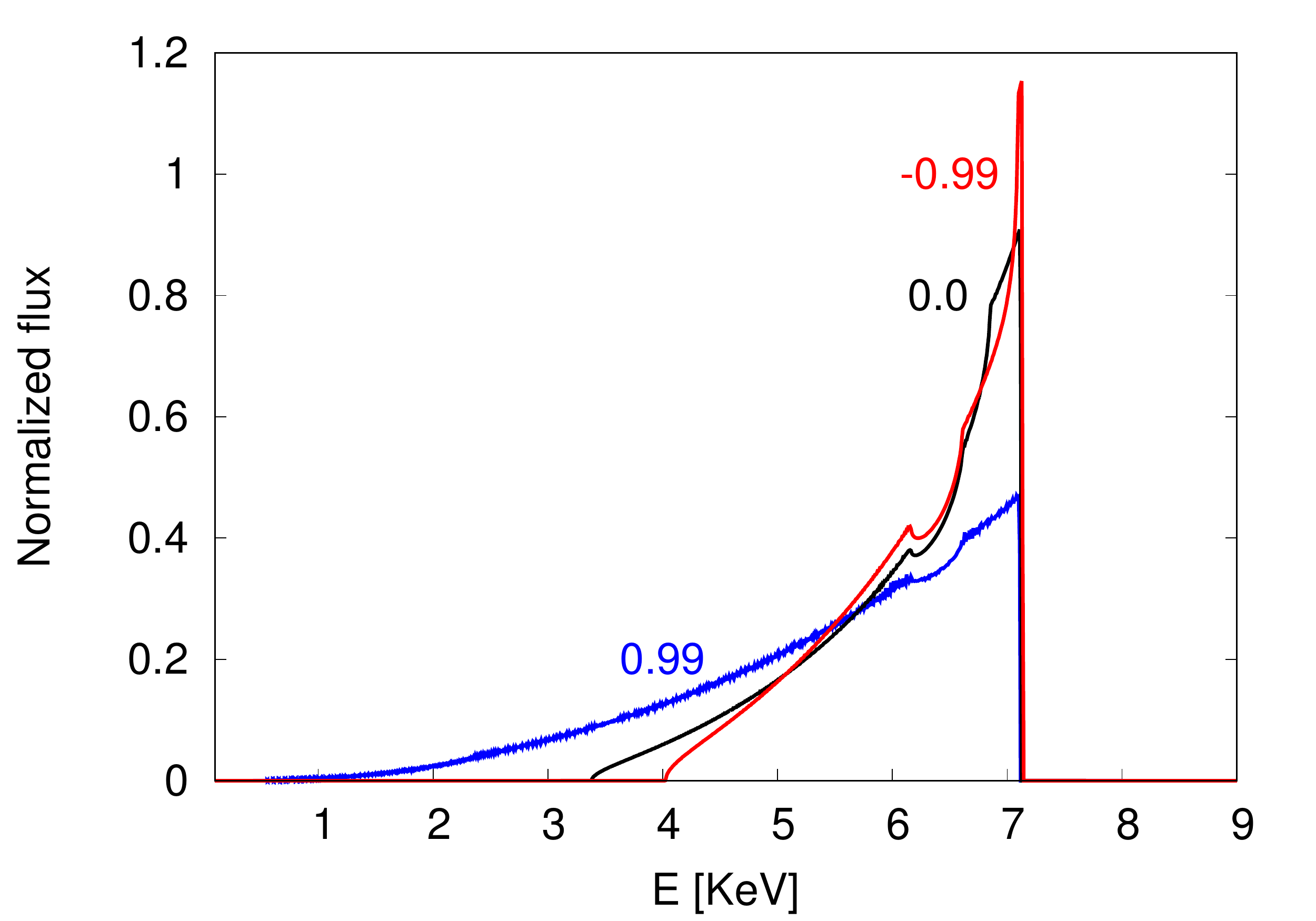} 
\hspace{0.5cm}
\includegraphics[type=pdf,ext=.pdf,read=.pdf,width=7cm]{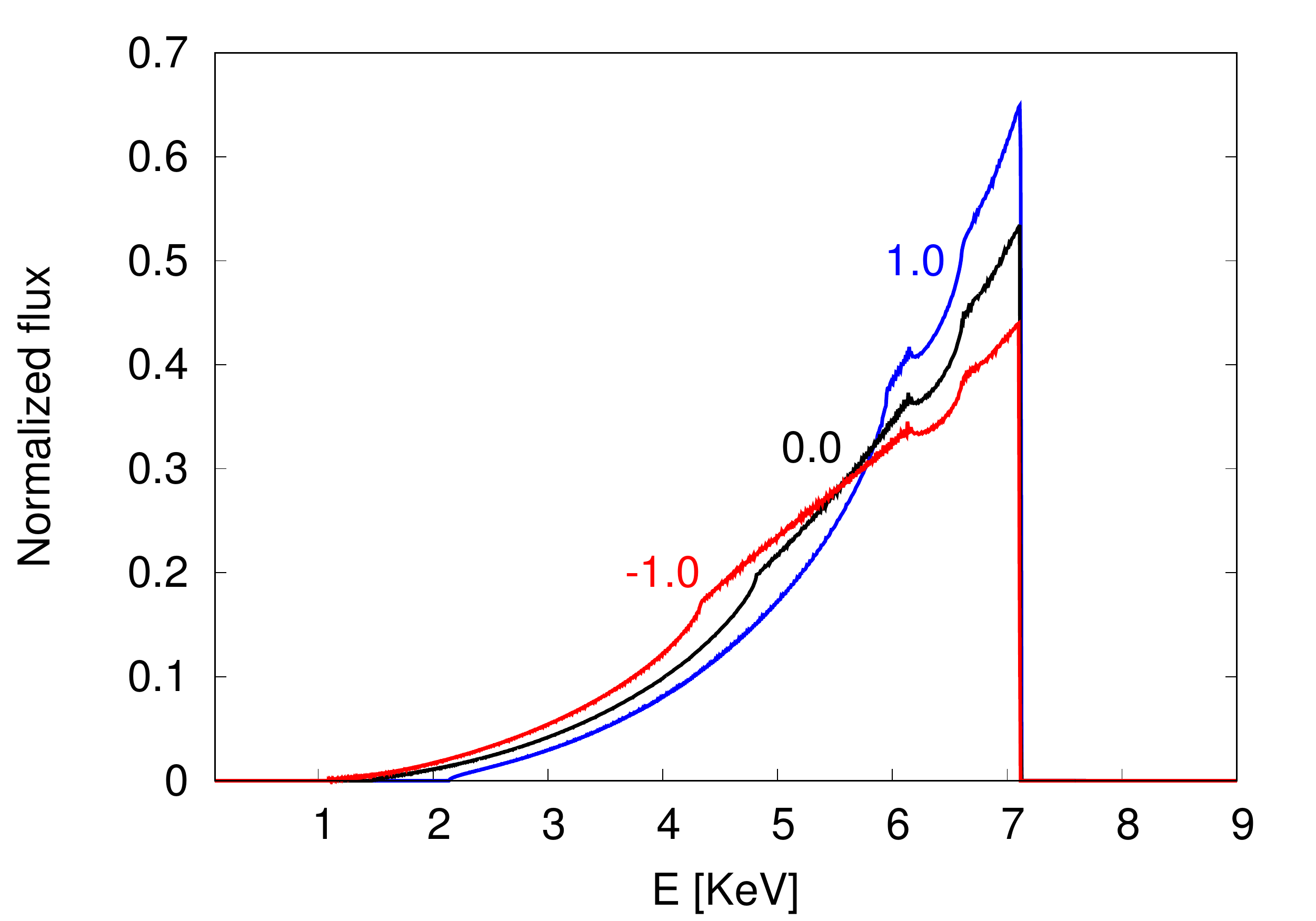}
\end{center}
\vspace{-0.4cm}
\caption{Normalized flux plotted as a function of energy at the observer using {\sc relline\_nk}. The left figure plots Kerr cases, with the labels indicating the value of $a_*$, while the right figure plots non-Kerr cases at a fixed $a_*=0.9$, with the labels denoting the value of $\alpha_{13}$. Inclination is fixed in both cases at $45^{\circ}$. \label{fig:ironlines}}
\end{figure*}

The reflection spectrum is a combination of radiative recombination continua, fluorescent lines (most notably the iron~K complex at $6\sim7$~keV), absorption edges, and a Compton hump at $20\sim30$~keV. This radiation carries information on the physical composition of the accretion disk and on the strong gravitational field around the black hole. The Fe~K emission line (and other fluorescent lines) are broadened and skewed by relativistic effects (Doppler effects, gravitational redshift, and light bending)~\cite{Fabian:1989ej,Laor:1991nc}. To illustrate the relativistic effect of the black hole, we show in Fig.~\ref{fig:ironlines} left panel the broadened iron lines for three representative values of the black hole spin, viz., $-0.99$, 0 and $0.99$, while keeping all deformations zero. The clear change in the shape of the line for different spin shows that  by fitting for the line, one can recover the spin of the black hole. This has been a standard technique for measuring black hole spin for a decade. In the right panel of the same figure, we show the lines, keeping spin constant, for three representative values of the Johannsen deformation parameter $\alpha_{13}$. The shape of the line also changes with $\alpha_{13}$, and therefore, reflection spectroscopy can also be used to estimate the deviation from Kerr (see Ref.~\cite{bambi2016} for the effect of other Johannsen deformation parameters on the iron line shape).

The reflection model most widely used in the past for both general application and measuring black hole spins is \textsc{reflionx}~\cite{Ross:2005dm}. Recently, a more sophisticated reflection model, \textsc{relxill}, has been developed. It is based on the reflection code \textsc{xillver}~\cite{garcia2011x,garcia2013x,garcia2014improved}, and the relativistic line-emission code \textsc{relline}~\cite{dauser2010broad,dauser2013irradiation,dauser2014role}. Compared to \textsc{reflionx}, \textsc{relxill} has several superiorities. For instance, it produces reflection with the most updated atomic database provided by the photoionization code \textsc{xstar}~\cite{kallman2001photoionization}, and thus delivers a superior calculation of the ionization balance.

By calculating the transfer function for a generic stationary, axisymmetric, and asymptotically flat black hole spacetime, an extended relativistic reflection model \textsc{relxill\_nk} has been constructed (see Refs.~\cite{bambi2016,abdikamalov2019} for details on the model and \cite{review-cp} for a review on our results). Presently, \textsc{relxill\_nk} can work either with the Johannsen metric to measure the deformation parameters $\alpha_{13}$, $\alpha_{22}$, and $\epsilon_{3}$~\cite{bambi2016,abdikamalov2019}, or with a black hole solution from a class of conformally invariant theories of gravity with the deformation parameter $L$~\cite{conf1,conf2}. The analysis in this paper is limited to $\alpha_{13}$, which has the largest impact on the shape of reflection spectrum among the deformation parameters of the Johannsen metric; this means that the other deformation parameters are set to be zero. The demonstration of the impact of other parameters is left to other work.


\section{Observations, data reduction, and combining spectra} \label{obs}

We use observations from the X-ray mission \textit{Rossi X-ray Timing Explorer (RXTE)} which has the principal detector the \textit{Proportional Counter Array (PCA)} on board containing five \textit{Proportional Counter Units (PCUs)}, with sensitivity from $2\sim60$~keV. Even though the spectral resolution of the instrument (17\% at 6~keV) is limited, the aggregate observing time and high-throughput signal of RXTE are so large that we could obtain spectra for many black hole systems, each with such high signal-to-noise that we are effectively systematics limited.

Compared to other X-ray instruments, RXTE/PCA has several advantages. Prominently, it is free from the problematic effects of pileup, which is commonly a serious problem in analyzing and interpreting data for bright sources obtained using CCD detectors, which occurs when several photons hit the detector at the same region between two read-outs, causing confusion in the signal from two distinct events combined as one with their energies summed. Another attribute of the PCA, which has only recently been surpassed by the Nuclear Spectroscopic Telescope Array (NuSTAR \cite{nustar}), is its high-energy sensitivity, which allows observations of both the Fe K region and the Compton hump using a single detector. By implementing a calibration tool called \textsc{pcacorr}, the sensitivity of the RXTE PCA detector to faint spectral features (such as the Fe~K line/edge) reaches $\sim0.1\%$~\cite{garcia2014_pcacorr}\footnote{After applying the calibration tool \textsc{pcacorr}, only 0.1\% systematic uncertainty is sufficient to achieve acceptable fit ($\chi^2_\nu \sim 1$), rather than the value of 0.5\%-1\% previously used for RXTE spectra.}.

We test our non-Kerr model on GX~339$-$4 using the observations previously analyzed in Ref.~\cite{javier_gx339}, because of the high signal-to-noise of the data and because we can then compare the results here with those in~\cite{javier_gx339}. For simplicity, here we only consider the brightest data from GX~339$-$4's hard state. We analyze just the data collected using the best-calibrated detector, PCU-2, which also provides the richest data set. In order to boost signal-to-noise, we produce a composite PCU-2 hard-state spectrum of GX~339$-$4 with a total exposure time of 46~ks, which is a summation of 23 individual exposures taken during the 2002 outburst, all of them corresponding to roughly the same source intensity ($L/L_{\rm Edd} = 0.17$ assuming a distance of 8~kpc and a black hole mass of 10~$M_\odot$). The total number of source counts is 40 million and the counts-per-keV in the continuum at 6.4~keV is 4.4~million, while the total number of counts in the Fe K line region ($3\sim10$~keV) is 28~million. Now, since we have a newer version of \textsc{relxill} package that could introduce possible deviation from the Kerr metric, we could look at this dataset from a new perspective.

The main reason for choosing an RXTE observation rather than NuSTAR or XMM-Newton data is that the luminosity of this particular RXTE spectrum is 17\%~$L_{\rm Edd}$ and is the highest for GX~339$-$4 in the low-hard state (just before transitioning into the high-soft state). While there is still debate on whether and when the disk becomes truncated in the low-hard state, it is agreed that, as the accretion luminosity increases, the inner edge of the disk moves to smaller radii, which makes the assumption $R_{\rm in}=R_{\rm ISCO}$ stronger and helps to break the complicated degeneracy among $a_*$, $\alpha_{13}$, and $R_{\rm in}$. Moreover, our previous endeavor with 6~joint observations with Swift and NuSTAR in the 2015 outburst of GX~339$-$4 leads to much worse constraints on both $a_*$ and $\alpha_{13}$ (while still recovering the Kerr metric).

All spectral fitting is done with one of the standard software systems for analyzing X-ray spectral data, {\sc Xspec}~\cite{arnaud}, and in the \textsc{Tbabs} model, we used the \textit{Wilm} set of interstellar medium abundances \cite{wilms2000absorption} and \textit{Verner} photoelectric cross sections \cite{a3}. The development of model and their statistics are shown in Tab.~\ref{tab:chisq_change} and will be discussed in the next section. All the uncertainties quoted in this paper are for a 90\% confidence range, unless otherwise stated.


\section{Spectral fitting and results}\label{fitting}

We firstly fit the data with an absorbed powerlaw (\textsc{Tbabs*powerlaw} in {\sc Xspec} language) with a fixed column density\footnote{Throughout our study, we use such a value for the column density, which is obtained from the HEASARC web tool nH available at
https://heasarc.gsfc.nasa.gov/cgi-bin/Tools/w3nh/w3nh.pl \cite{Bekhti:2016}.} $N_{\rm H} = 4\times10^{21}$ cm$^{-2}$, and the residuals show a clear signature of reflection including the Fe K line/edge and the Compton hump (see Fig.~\ref{fig:series_models}~(a)). Thus, a relativistic reflection model is fundamentally needed in the following fits. The raw data can be seen in the top panel in Fig.~\ref{f-datamodel}.

Besides the background metric, the spectral shape of reflection also depends on the geometry of the emitting region (the black hole spin $a_*$, the inner and the outer edge of the emission region $R_{\rm in}$ and $R_{\rm out}$), the disk properties (ionization state, Fe abundance), the inclination angle of the disk with respect to the line of sight ($i$), and the disk emissivity profile. The intensity profile is usually modeled as a power-law $I_e\propto r^{-q}$, where $q$ is called the emissivity index. In each of our model choices, we tried emissivity profiles including:
\begin{enumerate}[nosep]
  \item The canonical emissivity profile, $I_e\propto r^{-3}$ ($q=3$).
  \item Free emissivity index $q$ all over the disk, i.e., $I_e\propto r^{-q}$. This profile performs the best both in statistics and in parameter constraints in all models, so its results were chosen to present, unless otherwise stated. 
  \item Lamppost geometry, in which the corona is assumed to be a point source on the spin axis above the black hole source \cite{lamppost}. We note that the non-Kerr relativistic reflection spectrum in lamppost geometry was not available when this work was done, and we later found that the Kerr lamppost model does not result in a better fit than free power-law emissivity profile in all models. 
    \item In general, a simple power-law is not expected to provide a good description of the emissivity profile of accretion disks illuminated by X-ray coronae, and a broken power-law, or a twice-broken power-law, is necessary~\cite{Miniutti:2003yd,Wilkins:2011kt,Wilkins:2012zm}. However, when we fit our data of GX~339$-$4, we find that a broken power-law emissivity leads to comparable statistics as the emissivity of a simple power-law with $q$ free. Even in the simplest case in which we only have two free parameters, namely the inner emissivity index $q_1$ and the breaking radius $R_{\rm br}$ while the outer emissivity index is frozen to $q_2=3$ (Newtonian limit in the lamppost geometry at large radii), we are unable to constrain the breaking radius $R_{\rm br}$. For this reason, we prefer the simple power-law case. We stress, however, that the choice of a broken power-law leads to contours of $\alpha_{13}$ and $a_*$ very similar to the simple power-law case.
\end{enumerate}

We make the assumption that the inner disk is truncated at the ISCO, as explored previously for this data~\cite{javier_gx339}. This is the key to constraining the deformation parameter $\alpha_{13}$ and $a_*$. The reason is that $R_{\rm in}$ (set equal to $R_{\rm ISCO}$), $\alpha_{13}$ (which has a significant impact on $R_{\rm ISCO}$) and $a_*$ (which maps $R_{\rm ISCO}$ one to one for fixed $\alpha_{13}$), are all interconnected and thus have a strong degeneracy waiting to be broken.

\subsection{Model 1: relativistic reflection}\label{model1}

Under such assumptions, in both Kerr and non-Kerr fits, namely using \textsc{Tbabs*relxill} and \textsc{Tbabs*relxill\_nk} respectively, the residuals show that two absorption features flank the Fe K line at $\sim5.6$ and~$\sim7.2$~keV, and a Compton hump at higher energies is present (Fig.~\ref{fig:series_models}~(b)). After relaxing the Kerr requirement by introducing the deformation parameter $\alpha_{13}$, the fit with only the relativistic reflection was improved when $\chi^2$ reduced by 28 with one extra free parameter.

We have also tried to treat the incident spectrum as an \textsc{nthcomp} Comptonization continuum (\textsc{Cp}) instead of the standard high-energy cutoff powerlaw. With the model \textsc{tbabs*relxillCp\_nk}, a prominent feature around 7 keV is clearly visible as shown in Fig.~\ref{fig:series_models}~(f), whose residual shape is similar to that with model \textsc{Tbabs*relxill}. This feature would not vanish when we later tried to add a corresponding \textsc{xillverCp} component, but would vanish if we included an absorption via \textsc{gabs} (see Section~\ref{model3}). The best-fit parameter values are shown in Tab.~\ref{tab:parameters}. Notice that the model \textsc{tbabs*relxillCp\_nk*gabs} has found the energy of the absorption line is at $7.22^{+0.09}_{-0.06}$~keV, which is perfectly in alignment with the value that had been found in Ref.~\cite{javier_gx339} ($7.23\pm0.08$~keV).

As we can see from Tab.~\ref{tab:chisq_change}, the reduced $\chi^2$ is 3.8 or higher (depending on the choice of the emissivity profile) when we use \textsc{relxill} and \textsc{relxill\_nk}. It becomes 2.4 (for every choice of the emissivity profile) when we employ \textsc{relxillCp} and \textsc{relxillCp\_nk}.

\subsection{Model 2: both relativistic and unblurred reflections}\label{model2}

\begin{figure*}[t]
\includegraphics[type=pdf,ext=.pdf,read=.pdf,width=1.\textwidth]{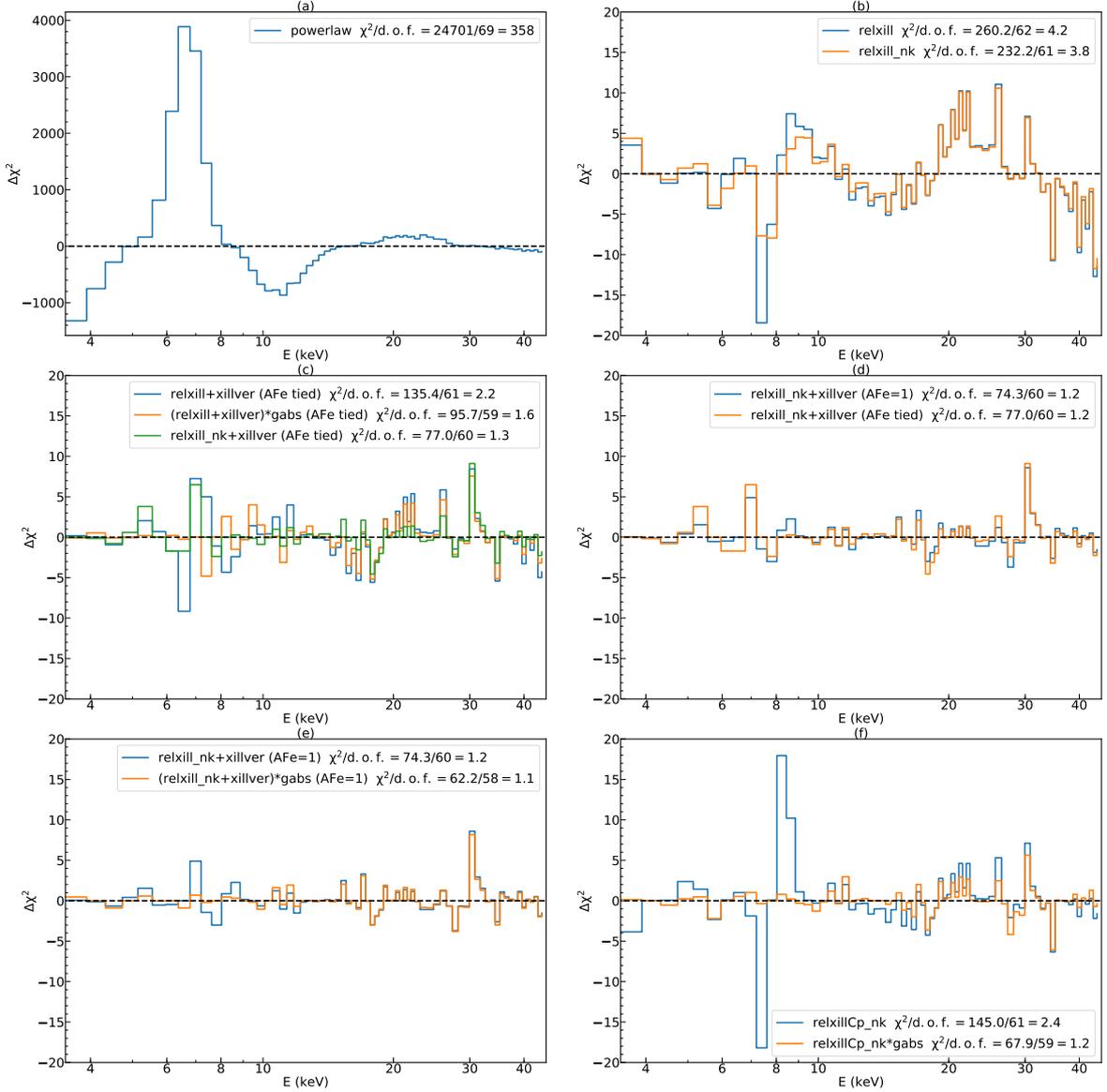}
\caption{Contributions to $\chi^2$ for the spectrum resulting from fitting. See the context for more details. \label{fig:series_models}}
\end{figure*}

The residuals in Fig.~\ref{fig:series_models}~(b) for the models \textsc{Tbabs*relxill} and \textsc{Tbabs*relxill\_nk} are significantly reduced by introducing an unblurred reflection component via \textsc{xillver} (i.e. \textsc{Tbabs*(relxill+xillver)} and \textsc{Tbabs*(relxill\_nk+xillver)}). In particular, the 5.6~keV ``absorption'' feature disappears after the unblurred reflection is added. If we consider the emissivity profile with free $q_1=q_2$, and with the non-Kerr metric, the introduction of an unblurred reflection component, to the fundamental relativistic reflection, has a F statistic value of at least 120.9 ($A_{\rm Fe}$ tied or not), which suggests this component is significant. The unblurred reflection could physically originate from cold material in a wind, the outer region of a flared disk, or the companion star. Thus, the ionization parameter in \textsc{xillver} was fixed at its minimum value, $\log\xi= 0$, while the other parameters were linked to those in the relativistic reflection.

In Ref.~\cite{javier_gx339}, the authors have found that the data required the unblurred reflection to have near-solar Fe abundance, while the inner-disk abundance was quite high ($A_{\rm Fe}\sim5$). A more plausible picture, using a single value of abundance, would lead to a much worse fit, which was the reason why they fixed the abundance to be solar in \textsc{xillver} ($A_{\rm Fe}=1$) in their final fits despite defying a natural explanation. In view of the study presented in~\cite{Jiang:2019xqn}, we can argue that the high iron abundance required in the relativistic component may be due to the higher disk electron density (here and in~\cite{javier_gx339} fixed to $n_{\rm e} = 10^{15}$~electrons/cm$^3$). Another possible explanation is that the effect of returning radiation, namely of the radiation emitted by the disk and returning the disk because of the strong light bending near the black hole (not included in \textsc{relxill} and \textsc{relxill\_nk}) may mimic a higher iron abundance in the disk~\cite{Ross:2002ix}.

In our fits, when the Fe abundances are tied in the relativistic and distant reflectors and $q$ is free, relaxing the Kerr requirement would reduce the $\chi^2$ by 58. The spin is found to be extremely high ($a_*=0.9861^{+0.0004}_{-0.0039}$) and the scaled deformation parameter $\alpha_{13}$ only has an upper limit of $-0.987$; that is, we do not recover the Kerr solution at a high confidence level (see Tab.~\ref{tab:parameters}). In addition, it performs even better than the \textsc{Tbabs*(relxill+xillver)*gabs} fit, especially around 20~keV (see Fig.~\ref{fig:series_models}~(c)).

Surprisingly, in the non-Kerr metric, with either treatment for Fe abundance in \textsc{xillver}, the fits are rather comparable in nearly all energy bins, as shown in Fig.~\ref{fig:series_models}~(d), and the best fit values are presented in Tab.~\ref{tab:parameters}. The fits for the models with tied iron abundance are much better than the Kerr case. Thus, the deformation parameter has shown its potential to settle the unexpected discrepancy of Fe abundance in inner-disk and distant reflections in previous study.

\begin{table*}
\begin{center}
{\tiny
\caption{Statistics ($\chi^2/d.o.f.$) of the individual fits to the spectrum. Note that the non-Kerr relativistic reflection spectrum in lamppost geometry was not available when this work was done. $^\dagger$ The models all include the galactic absorption model with a fixed column density $4.0\times 10^{21}$cm$^{-2}$.  In addition, $A_{\rm Fe}$ tied or $A_{\rm Fe}=1$ refers to the Fe abundance treatment in \textsc{xillver}, while in \textsc{relxill} $A_{\rm Fe}$ is always free.\label{tab:chisq_change}}
{\renewcommand{\arraystretch}{1.3}%
\begin{tabular}{ccccc}\hline \hline
\hspace{0.3cm} Model$^\dagger$ \hspace{0.3cm}&\hspace{0.3cm}$q_1=q_2=3$\hspace{0.3cm}& \hspace{0.3cm} $q_2=3$, free $q_1$ and $R_{br}$ \hspace{0.3cm}&\hspace{0.3cm}free $q_1=q_2$\hspace{0.3cm}&\hspace{0.3cm}lamppost (lp)\hspace{0.3cm}\\
\textsc{relxill} & 322.3/63=5.1 & 266.2/61=4.4& 260.2/62=4.2 & 307.4/63=4.9 \\
\textsc{relxill\_nk} & 315.3/62=5.1 & 233.3/60=3.9 & 232.2/61=3.8 &  \\
\hline
\textsc{relxill+xillver} ($A_{\rm Fe}$ tied)& 224.9/62=3.6 & 133.8/60=2.2& 135.4/61=2.2 & 210.9/62=3.4 \\
\textsc{relxill\_nk+xillver} ($A_{\rm Fe}$ tied) & 103.8/61=1.7 & 102.1/59=1.7 & 77.0/60=1.3 &  \\
\textsc{relxill+xillver} ($A_{\rm Fe}=1$)& 141.5/62=2.3 & 81.9/60=1.4& 82.0/61=1.4 & 133.4/62=2.2 \\
\textsc{relxill\_nk+xillver} ($A_{\rm Fe}=1$) & 77.2/61=1.3 & 74.0/59=1.2 & 74.3/60=1.2 &  \\
\hline
\textsc{(relxill+xillver)*gabs} ($A_{\rm Fe}$ tied)& 126.6/60=2.1 & 96.6/58=1.6& 95.7/59=1.6 & \\
\textsc{(relxill\_nk+xillver)*gabs} ($A_{\rm Fe}$ tied) & 103.1/59=1.7 & 95.6/57=1.6 & 77.0/58=1.3 &  \\
\textsc{(relxill+xillver)*gabs} ($A_{\rm Fe}=1$)& 67.2/60=1.1 & 62.3/58=1.1& 63.2/59=1.1 & \\
\textsc{(relxill\_nk+xillver)*gabs} ($A_{\rm Fe}=1$) & 62.8/59=1.1 & 62.3/57=1.1 & 62.2/58=1.1 &  \\
\hline
\textsc{relxillCp} & 152.8/63=2.4 & 146.7/61=2.4& 147.5/62=2.4 & 268.1/63=4.3 \\
\textsc{relxillCp\_nk} & 146.3/62=2.4 & 145.0/60=2.4 & 145.0/61=2.4 &  \\
\textsc{relxillCp*gabs} & 83.3/61=1.4 & 74.3/59=1.2& 74.3/60=1.2 & 149.5/61=2.4 \\
\textsc{relxillCp\_nk*gabs} & 78.8/60=1.3 & 66.5/58=1.1 & 67.9/59=1.2 &  \\
\hline 
\end{tabular}}
}
\end{center}
\end{table*}

\begin{table*}
\begin{center}
{\tiny
\caption{Best fit parameters for several targeted fits chosen from all of our test fits in Tab.~\ref{tab:chisq_change}, to investigate how the physical model choice and emissivity profile would make a difference to the resulting estimation of spin and deformation parameter. The fit labels (a) to (f) correspond to the contour plots in Fig.~\ref{fig:contours}. $^\dagger$ The models all include the galactic absorption model with fixed a column density $4.0\times 10^{21}$cm$^{-2}$. \label{tab:parameters}}
{\renewcommand{\arraystretch}{1.3}%
\begin{tabular}{cc|cc|cc|cc}\hline \hline
Model$^\dagger$ & Parameter &\multicolumn{2}{c|}{\textsc{relxill\_nk+xillver}} & \textsc{relxillCp\_nk} & \textsc{relxillCp\_nk*gabs} &\multicolumn{2}{c}{\textsc{(relxill\_nk+xillver)*gabs}} \\
&& (a) & (b) &(c)&(d)&(e)&(f)\\
\hline
\textsc{relxill(cp)\_nk}&$q$& $2.84\pm0.03$ & $2.894^{+0.007}_{-0.010}$ & $3.2^{+0.3}_{-0.2}$ & $2.48^{+0.02}_{-0.06}$ & $3^*$ & $2.8^{+0.4}_{-0.2}$  \\
\textsc{relxill(cp)\_nk}&$a_*$& $0.988^{+0.002}_{-0.005}$ & $0.9861^{+0.0004}_{-0.0039}$  & $-0.68^{+0.55}_{-0.19}$ & $0.988^{+0.003}_{-0.034}$ & $0.92^{+0.07}_{-0.12}$ & $0.987^{+0.007}_{-0.154}$ \\
\textsc{relxill(cp)\_nk}&$i$ (degrees)& $46.0^{+2.6}_{-0.2}$ & $48.4^{+0.2}_{-0.1}$ & $35.5^{+1.8}_{-0.9}$ & $40.0^{+1.3}_{-1.0}$ & $49\pm3$ & $51^{+3}_{-2}$ \\
\textsc{relxill(cp)\_nk}&$\Gamma$& $1.575^{+0.004}_{-0.006}$ & $1.544^{+0.007}_{-0.004}$ & $1.691^{+0.005}_{-0.004}$ & $1.671^{+0.005}_{-0.001}$ & $1.575^{+0.014}_{-0.013}$ & $1.580^{+0.010}_{-0.009}$ \\
\textsc{relxill(cp)\_nk}&$\log \xi$ (erg$\cdot$cm$\cdot$s$^{-1}$) & $3.47^{+0.04}_{-0.03}$ & $3.486\pm0.016$ & $3.35\pm0.03$ & $3.40^{+0.02}_{-0.04}$ & $3.37^{+0.07}_{-0.09}$ & $3.38^{+0.05}_{-0.06}$ \\
\textsc{relxill(cp)\_nk}&$A_{\rm Fe}$& $9.5^{+0.1}_{-0.2}$ & $3.14^{+0.02}_{-0.04}$ & $6.8^{+0.8}_{-0.9}$ & $>9.4$ & $7.7^{+2.0}_{-2.5}$ & $>5.3$ \\
\textsc{relxill\_nk}&$E_{\rm cut}$ (keV)& $85.4^{+1.2}_{-1.0}$ & $99.6^{+2.6}_{-1.1}$ & - & - & $86\pm3$ & $85^{+3}_{-2}$ \\
\textsc{relxillCp\_nk}&$kT_{\rm e}$ (keV)& - & - & $17.8\pm0.3$ & $16.4^{+0.2}_{-0.1}$ & - & - \\
\textsc{relxill(cp)\_nk}&$R_{\rm f}$& $0.251^{+0.003}_{-0.030}$ & $0.85^{+0.02}_{-0.04}$ & $0.21\pm0.01$ & $0.248^{+0.007}_{-0.013}$ & $0.29^{+0.04}_{-0.03}$ & $0.29^{+0.03}_{-0.04}$ \\
\textsc{relxill(cp)\_nk}&$\alpha_{13}$$^\dagger$ & $<-0.81$ & $<-0.987$ & $<0.13$ & $<-0.73$ & $-0.76^{+0.78}_{-0.60}$  & $<0.10$  \\
\textsc{relxill(cp)\_nk}&$N_{\rm relxill\_nk}$($10^{-3}$)& $35.0\pm0.5$ & $24.4\pm0.1$ & $34.7^{+0.4}_{-0.2}$ & $36.3^{+0.4}_{-0.3}$ & $35.4^{+0.9}_{-0.6}$ & $35.2\pm0.6$ \\
\textsc{xillver}&$A_{\rm Fe}$& $1^*$ & tied & - & - & $1^*$ & $1^*$ \\
\textsc{xillver}&$N_{\rm xillver}$($10^{-3}$)& $9.2^{+0.2}_{-0.9}$ & $6.4^{+0.2}_{-0.4}$ & - & - & $8.3^{+1.9}_{-1.6}$ & $9.8^{+2.0}_{-1.7}$ \\
\textsc{gabs}&$E_{\rm line}$ (keV)& - & - & - & $7.22^{+0.09}_{-0.06}$ & $7.40\pm0.14$ & $7.45^{+0.08}_{-0.11}$ \\
\textsc{gabs}&Stength& - & - & - & $0.056^{+0.016}_{-0.008}$ & $0.030^{+0.013}_{-0.009}$ & $0.029^{+0.008}_{-0.009}$ \\
\hline
&$\chi^2/d.o.f.$& $74.3/60=1.2$ & $77.0/60=1.3$ & $145.0/61=2.4$ & $67.9/59=1.2$ & $62.8/59=1.1$ & $62.2/58=1.1$ \\
\hline
\end{tabular}}\\
\vspace{0.3cm}
}
\raggedright{
\textbf{Notes.} Errors are at 90\% confidence level, are statistical only. Note that errors for model (e) are calculated for each parameter from the MCMC after marginalising over all other parameters. $^*$ indicates the parameter is frozen. $^\dagger$ The deformation parameter $\alpha_{13}$ is scaled in \textsc{relxill\_nk} to be in the range (-1,1), and the value and uncertainty of it are only unscaled to be the real one (depends on $a_*$) in the metric in (e) where we performed MCMC simulation, i.e., for all the other fits, the scaled values are shown. }
\end{center}
\end{table*}

\subsection{Model 3: both relativistic and unblurred reflections, with absorption}\label{model3}

In Ref.~\cite{javier_gx339}, the model \textsc{Tbabs*(relxill+xillver)} still has a strong residual feature at $\sim7.2$~keV that needs to be diminished by including a phenomenological Gaussian absorption component \textsc{gabs}, which was plausibly attributed to a highly ionized wind around the black hole or else an instrumental calibration defect.

The model \textsc{tbabs*(relxill\_nk+xillver)} ($A_{\rm Fe}$ tied) does not require an absorption feature, as attempting to add an absorption via \textsc{gabs} does not improve the fit (see Tab.~\ref{tab:chisq_change}).

\begin{figure}[t]
\begin{center}
\includegraphics[type=pdf,ext=.pdf,read=.pdf,width=8.5cm]{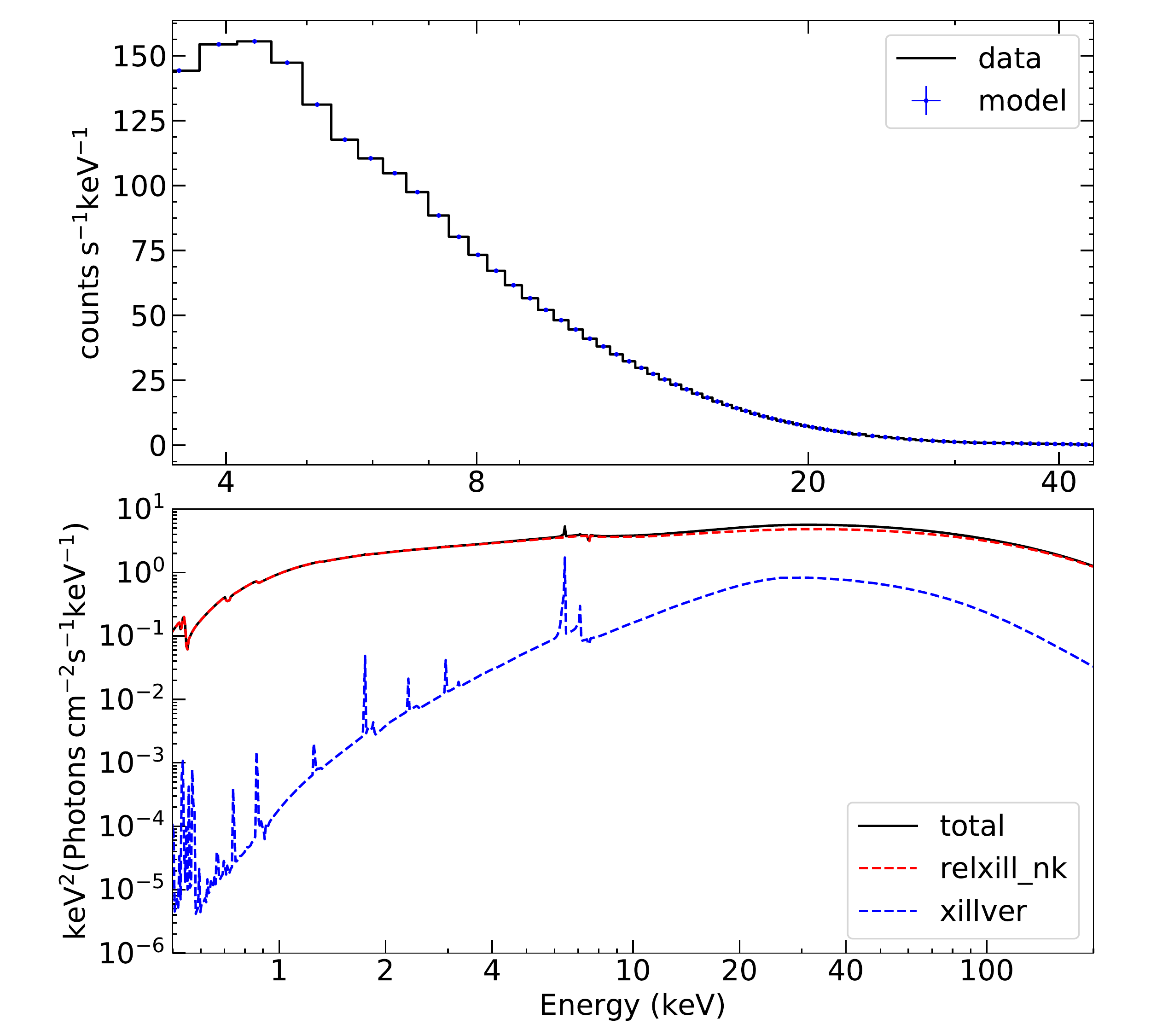}
\end{center}
\vspace{-0.1cm}
\caption{Data and the best-fit model \textsc{tbabs*(relxill\_nk+xillver)*gabs} with $A_{\rm Fe}=1$ in \textsc{xillver} and $q=3$ (top panel), and total spectrum with different model components (bottom panel). This is the fit we further performed an explorative MCMC calculation. \label{f-datamodel}}
\end{figure}

In contrast, when $A_{\rm Fe}=1$ in \textsc{xillver}, we found an absorption at $\sim7.4$~keV would further reduce $\chi^2$ by 13 with 2 more d.o.f. (see Fig.~\ref{fig:series_models}~(e)). This is a marginal improvement, but produces the best statistical fit, with $\chi^2/d.o.f.=62/58=1.07$. Fig.~\ref{f-datamodel} shows the data and the best-fit model (top panel) as well as the different model components (bottom panel). When comparing this fit and the fit with frozen emissivity index $q=3$, namely fit (e) and (f) in Tab.~\ref{tab:parameters}, we found they are quite similar regarding the best-fit values and the statistics, expect for that the free $q$ fit has a looser constraint on the Fe abundance and the deformation parameter. 


As seen in Tab.~\ref{tab:parameters}, we also note that the inclination determination is stable across fits, and falls into the range of $37^\circ<i<78^\circ$ from optical analysis \cite{optical}. 

For an exploratory purpose, on the fit \textsc{tbabs*(relxill\_nk+xillver)*gabs} ($A_{\rm Fe}=1$ in \textsc{xillver}, frozen $q=3$), we performed Markov Chain Monte-Carlo (MCMC) analysis using the \textsc{emcee-hammer} Python package \cite{foreman} which implements affine-invariant sampling via the {\sc Xspec} implementation described in~\cite{lmcx1}. With MCMC's great power in high-dimensional analysis, we can explore efficiently the parameter space, and directly determine the posterior probabilities for all the model parameters, calculate the confidence contours  and search for degeneracies. We utilize 80 walkers, which are initialized in a cluster distributed about the best fit we found in the model. Each walker has 870,000 elements in total, and the initial 400,000 elements are discarded as ``burn-in" period during which the chain reaches its stationary state. As the autocorrelation length is typically several thousand elements, the net number of independent samples in the parameter space we have is order of $10^4$. 

\begin{figure*}
\begin{center}
\includegraphics[type=pdf,ext=.pdf,read=.pdf,width=15cm]{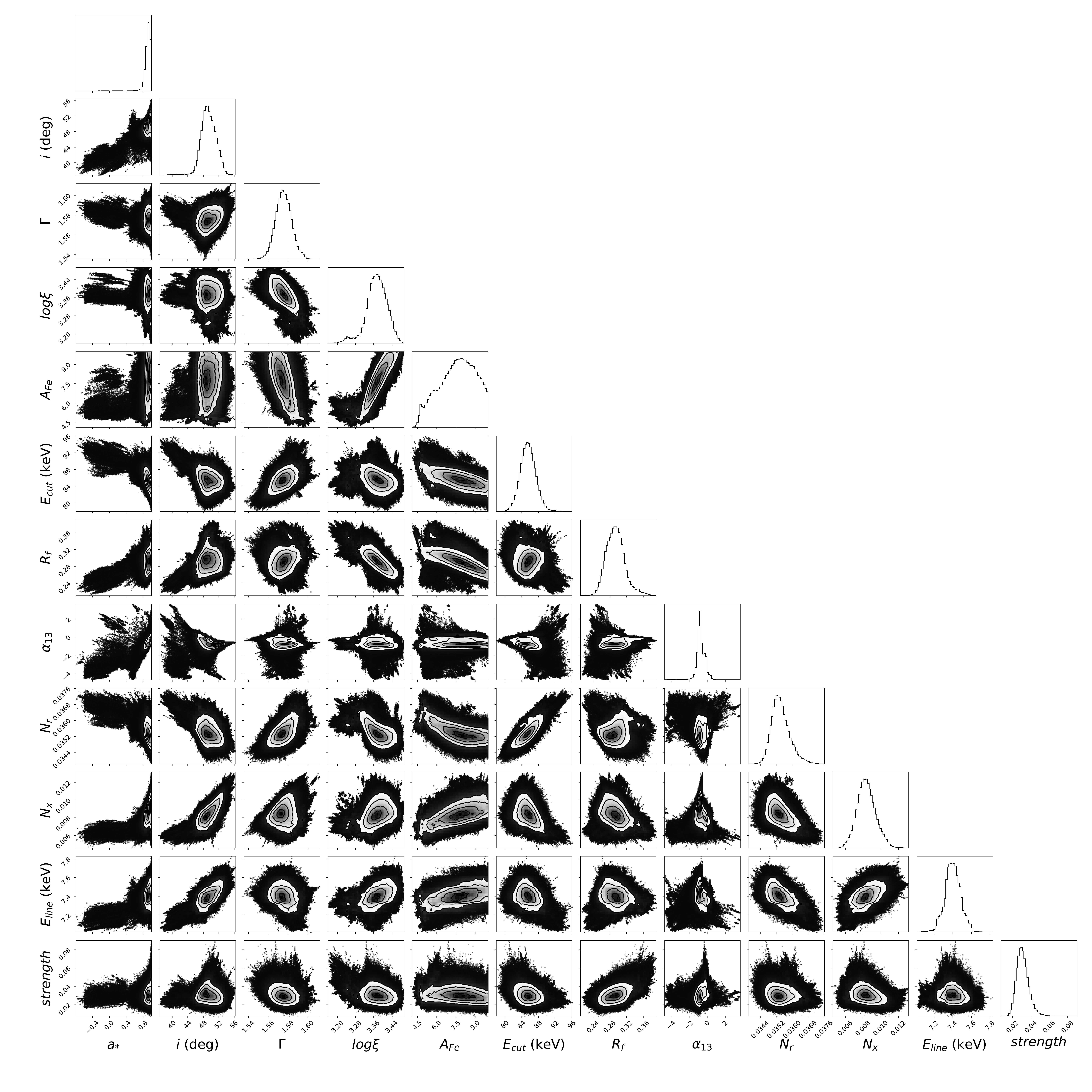}
\end{center}
\vspace{-0.8cm}
\caption{Corner plot for all the free parameter-pairs in Model \textsc{tbabs*(relxill\_nk+xillver)*gabs} ($A_{\rm Fe}=1$ in \textsc{xillver}, frozen $q=3$) after the MCMC run. \label{fig:corner}}
\end{figure*}

We show a corner plot with all the one and two dimensional projections of the posterior probability distributions of all free parameters in Fig.~\ref{fig:corner}. The contour plots are generated from the MCMC chains themselves (not from \textsc{steppar}). From the same figure we can also see which parameters are correlated and which are not. In particular, the deformation parameter $\alpha_{13}$ is correlated with the black hole spin $a_*$ and the disk's inclination angle $i$, but not much with the other parameters. In Table~\ref{tab:parameters}, we show the best fit values with uncertainties for all parameters, based on the marginalized distributions we obtain from the MCMC run. Our results are largely consistent with those found in~\cite{javier_gx339} under the assumption of the Kerr background.

\begin{figure}[t]
\begin{center}
\includegraphics[type=pdf,ext=.pdf,read=.pdf,width=8.5cm]{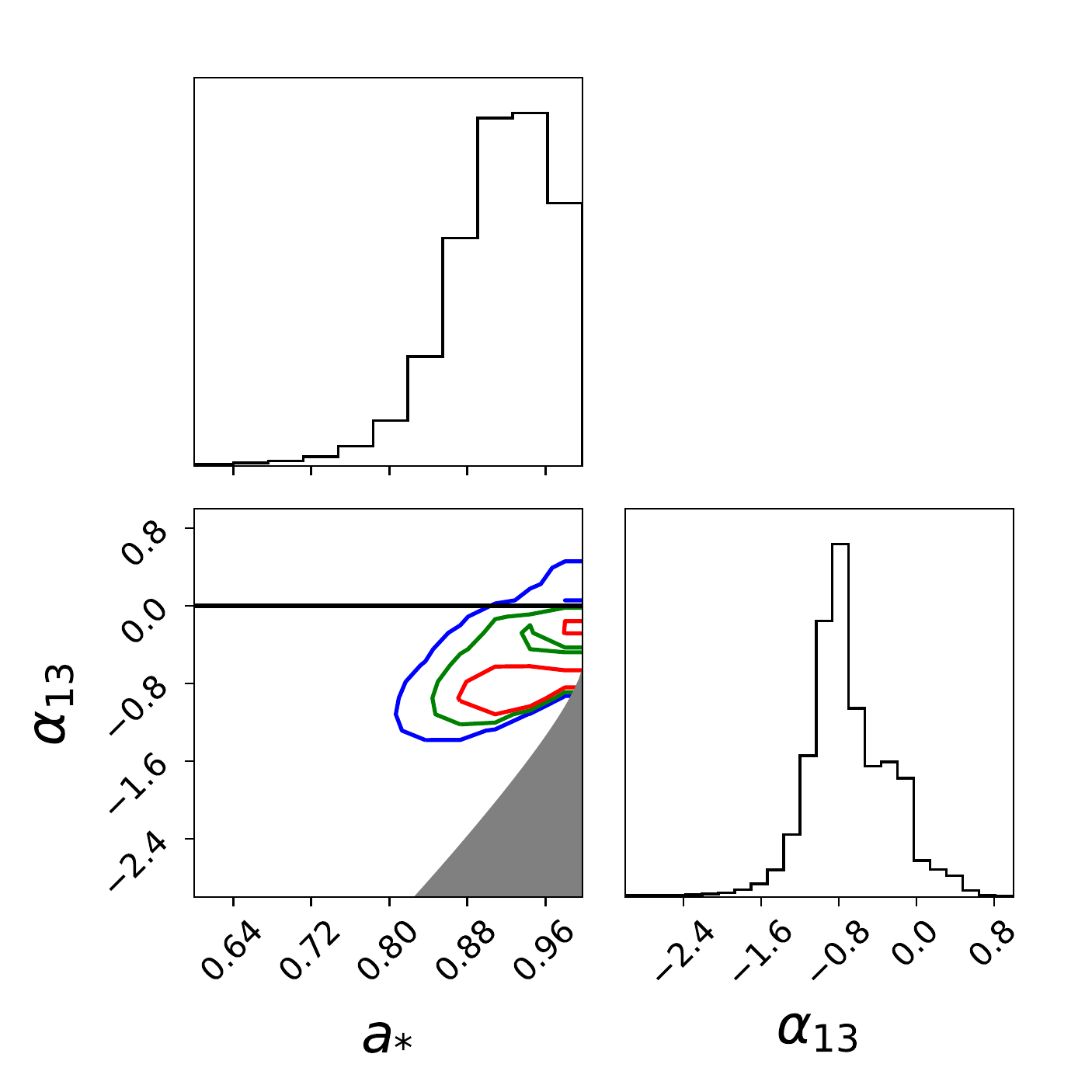}
\end{center}
\vspace{-0.5cm}
\caption{1-, 2-, 3-$\sigma$ confidence contours for $a_*$ and $\alpha_{13}$ in Model \textsc{tbabs*(relxill\_nk+xillver)*gabs} ($A_{\rm Fe}=1$ in \textsc{xillver}, frozen $q=3$) after the MCMC run. The grayed region is ignored in our study because it does not meet the condition in Eq.~\ref{eq:boundary-2}. \label{fig:contour}}
\end{figure}

In Fig.~\ref{fig:contour} we zoom into the $a_*$ vs. $\alpha_{13}$ part of Fig.~\ref{fig:corner}. Several features can be observed here: Firstly, there is some  degeneracy between the spin and the deformation parameter, as expected (see Sec.~\ref{reflection} for a discussion and Fig.~\ref{fig:ironlines}). In particular, within 90\% confidence $a_*=0.92^{+0.07}_{-0.12}$ and $\alpha_{13}=-0.76^{+0.78}_{-0.60}$. Secondly, this result is consistent with previous constraints on $\alpha_{13}$, as reported in~\cite{bh5,ashutosh}. Thirdly, the Kerr solution ($\alpha_{13}=0$, indicated by a solid black line in the figure) is recovered within 90\% confidence. Finally, we notice that the contours of degeneracy obtained here with an MCMC analysis are comparable to those obtained in~\cite{bh5,ashutosh}. This does not imply that there are no lingering systematic effects with these modeling efforts, an issue that will be examined further in a separate work.

\section{Discussion and Conclusions}\label{conclusion}

There are factors that have already been shown to play an important role in the estimation of the deformation parameter: the choice of physical model and the emissivity profile \citep{xu2018study, zhang2019kerr,liu19nk,zhang19nk}. In our case, this means that the treatment of iron abundance in \textsc{xillver}, the emissivity profile, whether to include absorption via \textsc{gabs}, and whether to see the incident spectrum as the cutoff powerlaw or an \textsc{nthcomp} Comptonization continuum, could all have an effect on the constraint of $\alpha_{13}$, and even the spin parameter. It is thus crucial to select the right model before testing GR. In principle, such a selection can be done by evaluating the quality of the fits from different models. However, depending on the quality of the data and the spectrum of the source, this can be challenging, as we find with the RXTE data of GX~339--4 analyzed in the present manuscript.

\begin{figure*}[t]
	\centering
	\subfloat[\textsc{relxill\_nk+xillver}, $A_{\rm Fe}$=1, free $q$, $\chi^2/d.o.f.=74.3/60=1.2$]{\label{figur:1}\includegraphics[type=pdf,ext=.pdf,read=.pdf,width=.45\textwidth,angle=-0]{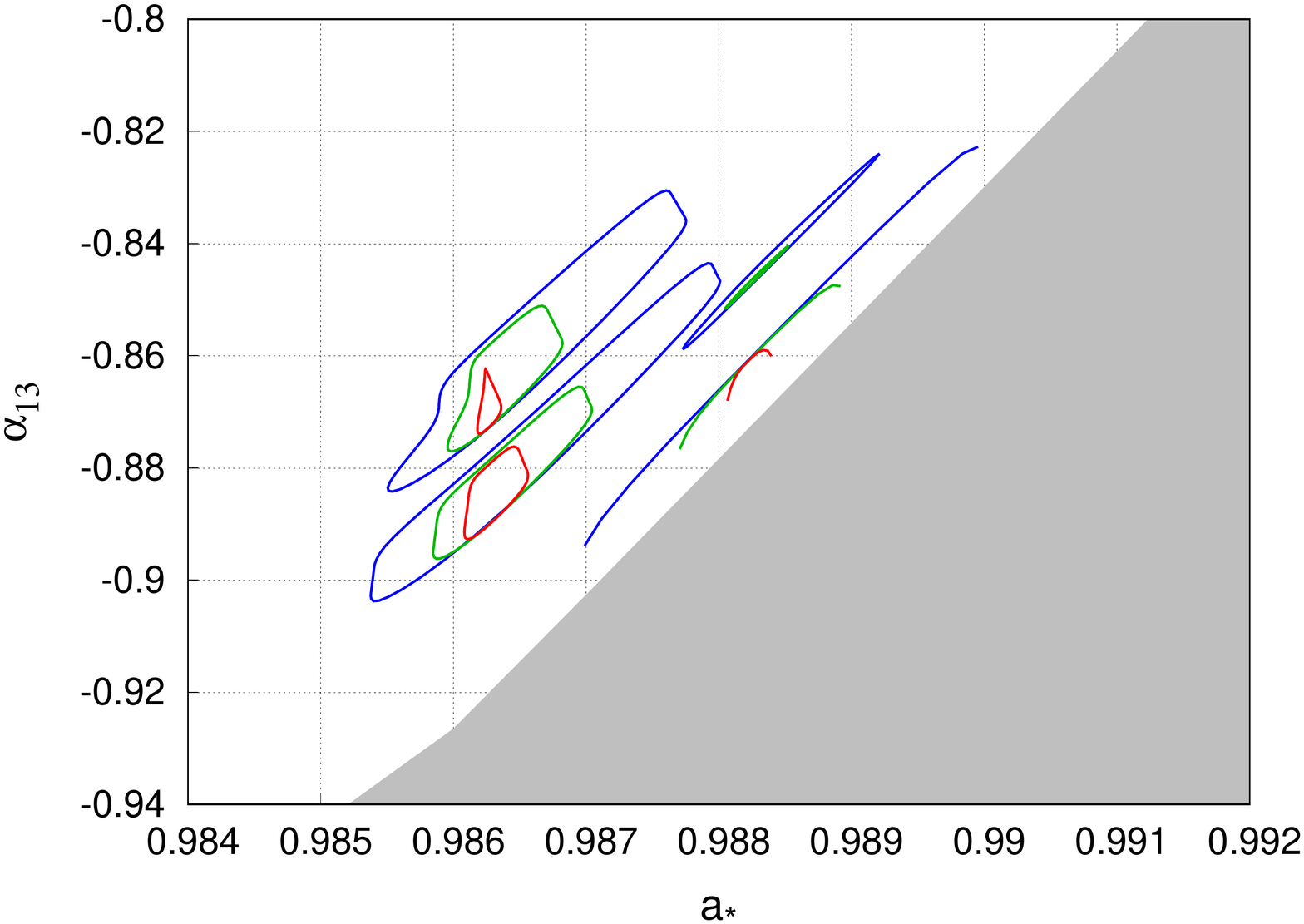}}\hfill
	\subfloat[\textsc{relxill\_nk+xillver}, $A_{\rm Fe}$ tied, free $q$, $\chi^2/d.o.f.=77.0/60=1.2$]{\label{figur:2}\includegraphics[type=pdf,ext=.pdf,read=.pdf,width=.45\textwidth,angle=-0]{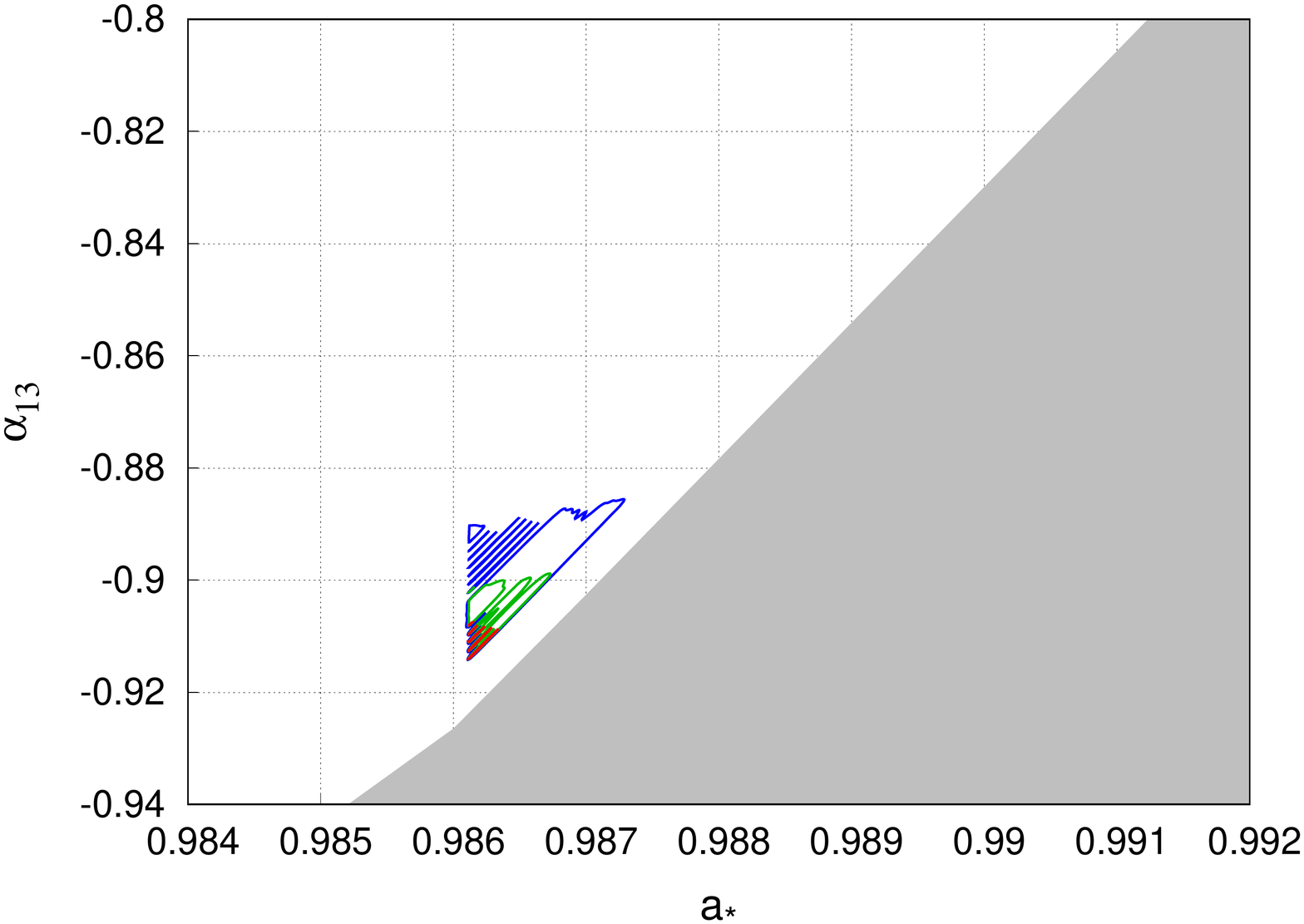}}
	\\
	\subfloat[\textsc{relxillCp\_nk}, free $q$, $\chi^2/d.o.f.=145.0/61=2.4$]{\label{figur:3}\includegraphics[type=pdf,ext=.pdf,read=.pdf,width=.35\textwidth,angle=-90]{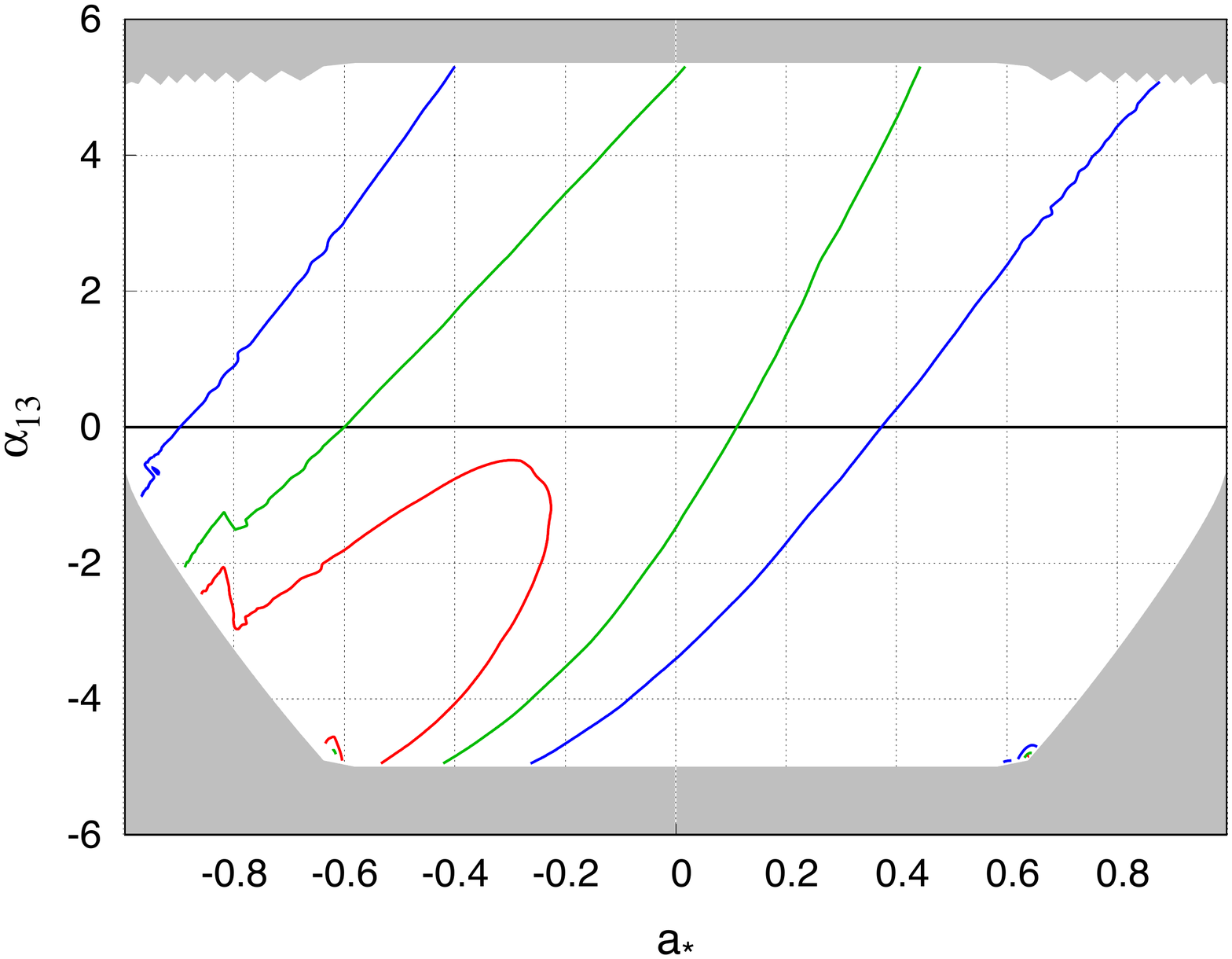}}\hfill
	\subfloat[\textsc{relxillCp\_nk*gabs}, free $q$, $\chi^2/d.o.f.=67.9/59=1.2$]{\label{figur:4}\includegraphics[type=pdf,ext=.pdf,read=.pdf,width=.35\textwidth,angle=-90]{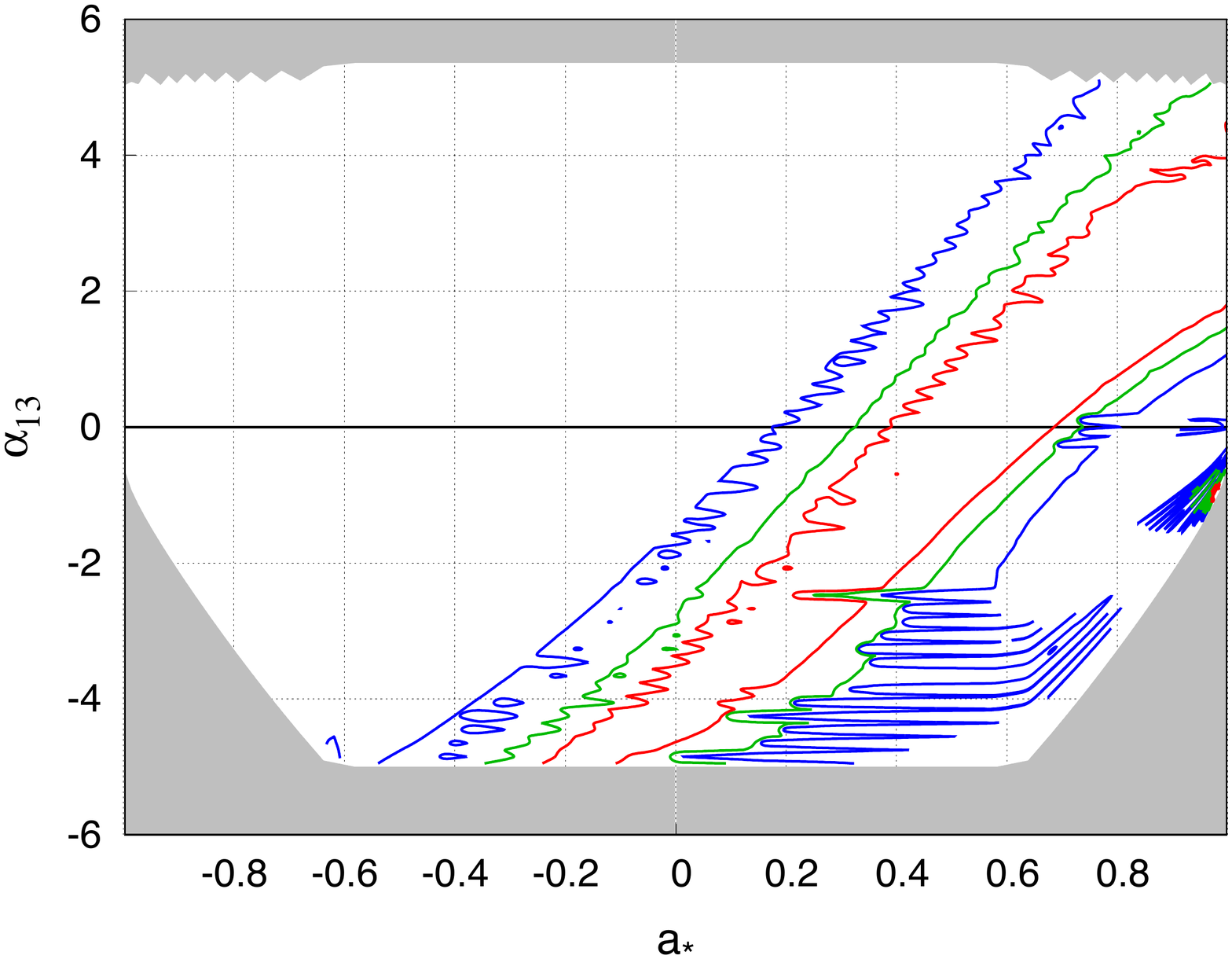}}
	\\
	\subfloat[\textsc{(relxill\_nk+xillver)*gabs}, $A_{\rm Fe}$=1, $q=3$, $\chi^2/d.o.f.=62.8/59=1.1$]{\label{figur:5}\includegraphics[type=pdf,ext=.pdf,read=.pdf,width=.35\textwidth,angle=-90]{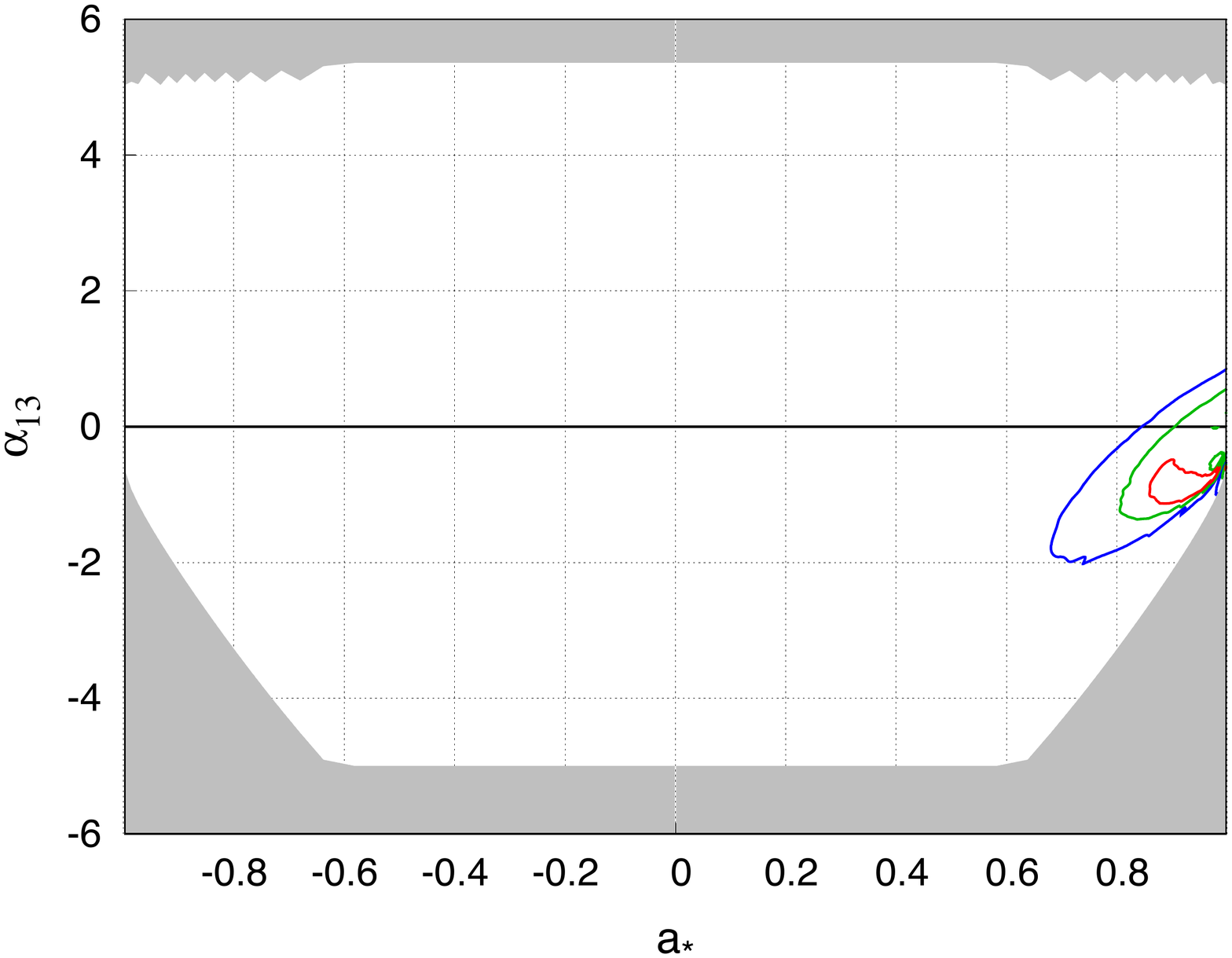}}\hfill
	\subfloat[\textsc{(relxill\_nk+xillver)*gabs}, $A_{\rm Fe}$=1, free $q$, $\chi^2/d.o.f.=62.2/58=1.1$]{\label{figur:6}\includegraphics[type=pdf,ext=.pdf,read=.pdf,width=.35\textwidth,angle=-90]{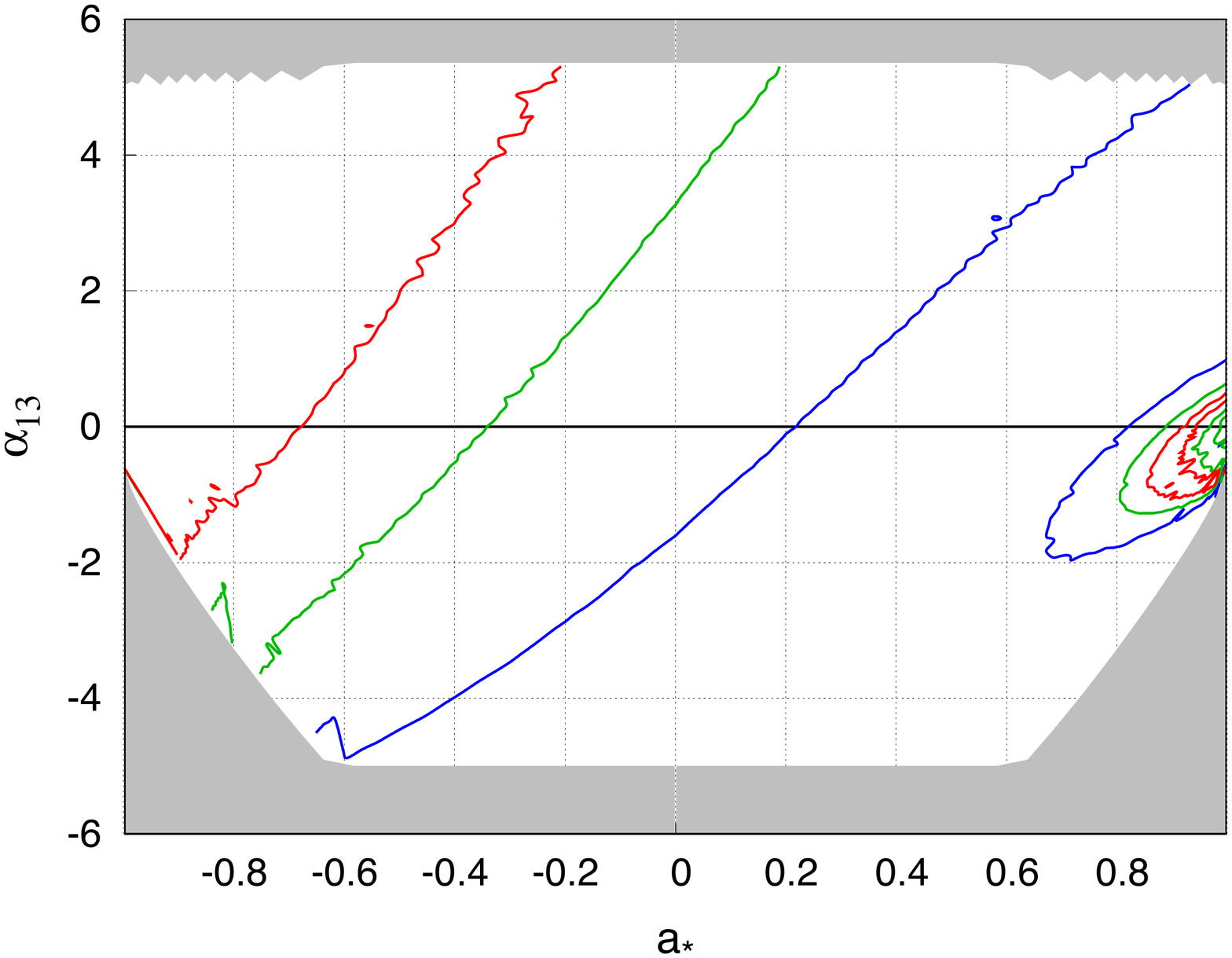}}
	\caption{1-, 2-, 3-$\sigma$ confidence contours for $a_*$ and $\alpha_{13}$ (scaled to be the real value in the Johannsen metric) with Model (a) to (f) in Tab.~\ref{tab:parameters} using \textsc{steppar}. The grayed region is ignored in our study because it does not meet the condition in Eq.~\ref{eq:boundary-2}. \label{fig:contours}}
	\end{figure*}

To investigate this, we have chosen 6 target fits from Tab.~\ref{tab:chisq_change}, presented the best-fit parameter values of them in Tab.~\ref{tab:parameters}. In {\sc Xspec}, we performed \textsc{steppar} calculations which generate the $\chi^2$ values stepping through given parameter ranges, in the $a_*$ vs. $\alpha_{13}$ parameter plane, and we show the resulting 1-, 2-, 3-$\sigma$ confidence contours in Fig.~\ref{fig:contours}.

The first message conveyed by these contour plots is that the choice of the physical model has significant impact in the resultant constraints on the deformation parameter. With the model \textsc{relxill\_nk+xillver}, the contour shape is consistent with the uncertainties we found in Tab.~\ref{tab:parameters}~(a) and (b), as the constraint on $a_*$ is very strong, and the scaled deformation parameter only has an upper limit of $-0.81$ and $-0.987$ when the Fe abundance is fixed to be solar and tied to the one in \textsc{relxill\_nk}. However, when the \textsc{Cp} flavor is adopted (Fig.~\ref{fig:contours}~(c) and (d)), the parameter space within the confidence contours becomes much wider, and its shape suggests a strong correlation between the spin and the deformation parameter which is as expected (see Fig.~\ref{fig:ironlines}). Comparing Fig.~\ref{fig:contours}~(c) and (d), we could see the elongate region within the contour lines moves to higher spin when an absorption is included, and a small region with very high spin and small ($\sim-1$) deformation parameter shows up which is overlapping with the one in (e) and (f).

\begin{figure}[t]
\begin{center}
\includegraphics[type=pdf,ext=.pdf,read=.pdf,width=0.5\textwidth]{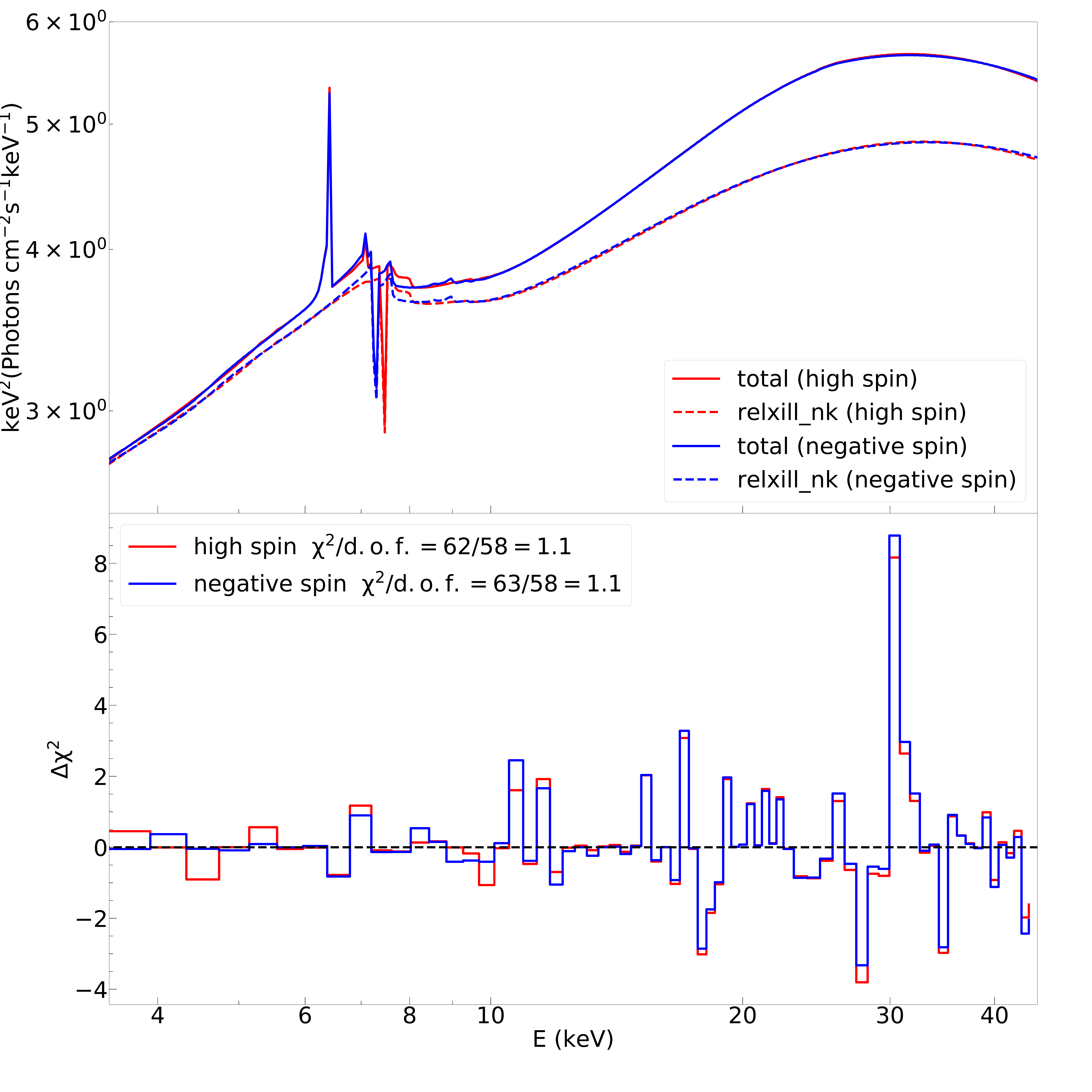}
\end{center}
\vspace{-0.5cm}
\caption{The predicted models and the distributions to $\chi^2$ in both statistically minima in Fig.~\ref{fig:contours}~(f), one at high spin and near-to-zero deformation parameter, and the other at negative spin and high value of deformation parameter. \label{fig:weird}}
\end{figure}

The contour in Fig.~\ref{fig:contours}~(e) obtained from \textsc{steppar} is consistent with our result from MCMC calculation (Fig.~\ref{fig:contour}). Since we also want to explore whether and how emissivity profile would change the constraint, we show the contour in Fig.~\ref{fig:contours}(f) with the same model but a free emissivity index $q$. The region at high spin and near-to-zero deformation parameter stays quite similar, but we find it surprisingly opens up a new region with negative spin and high value of deformation parameter. To investigate how this could happen, we plot the predicted models in both minima (with high spin and the negative spin) and the distributions to $\chi^2$ (see Fig.~\ref{fig:weird}). The largest discrepancy between these two minima is that they have found absorptions at slightly different energies ($\sim7.28$~keV for negative spin, $\sim 7.45$~keV for high spin), but the energy resolution of RXTE can hardly tell those apart, despite the extreme statistical precision. We have also compared the negative-spin ($a_*<-0.05$) to the high-spin fit in terms of other parameter values: the inclination is reduced from 51 deg to 40 deg, and the emissivity index is increased from $2.8^{+0.4}_{-0.2}$ to $6.4^{+1.9}_{-2.3}$, while the others stay similar; note that the higher emissivity index is unlikely because with a negative-spin, the corresponding ISCO radius increases, and with the assumption of $R_{\rm in}=R_{\rm ISCO}$, the actual scale of the disk's inner radius increases, making it implausible to have a large emissivity index. If the corona does not illuminate the ISCO region well, a fast-rotating black hole can be interpreted as a black hole with a low or negative value of the spin parameter~\cite{Fabian:2014tda}. We could thus use physically motivated priors to further constrain these models and select the right solution~\cite{Wilkins:2012zm,Fabian:2012kv,lamppost}.

\begin{figure}[t]
\begin{center}
\includegraphics[type=pdf,ext=.pdf,read=.pdf,width=0.5\textwidth]{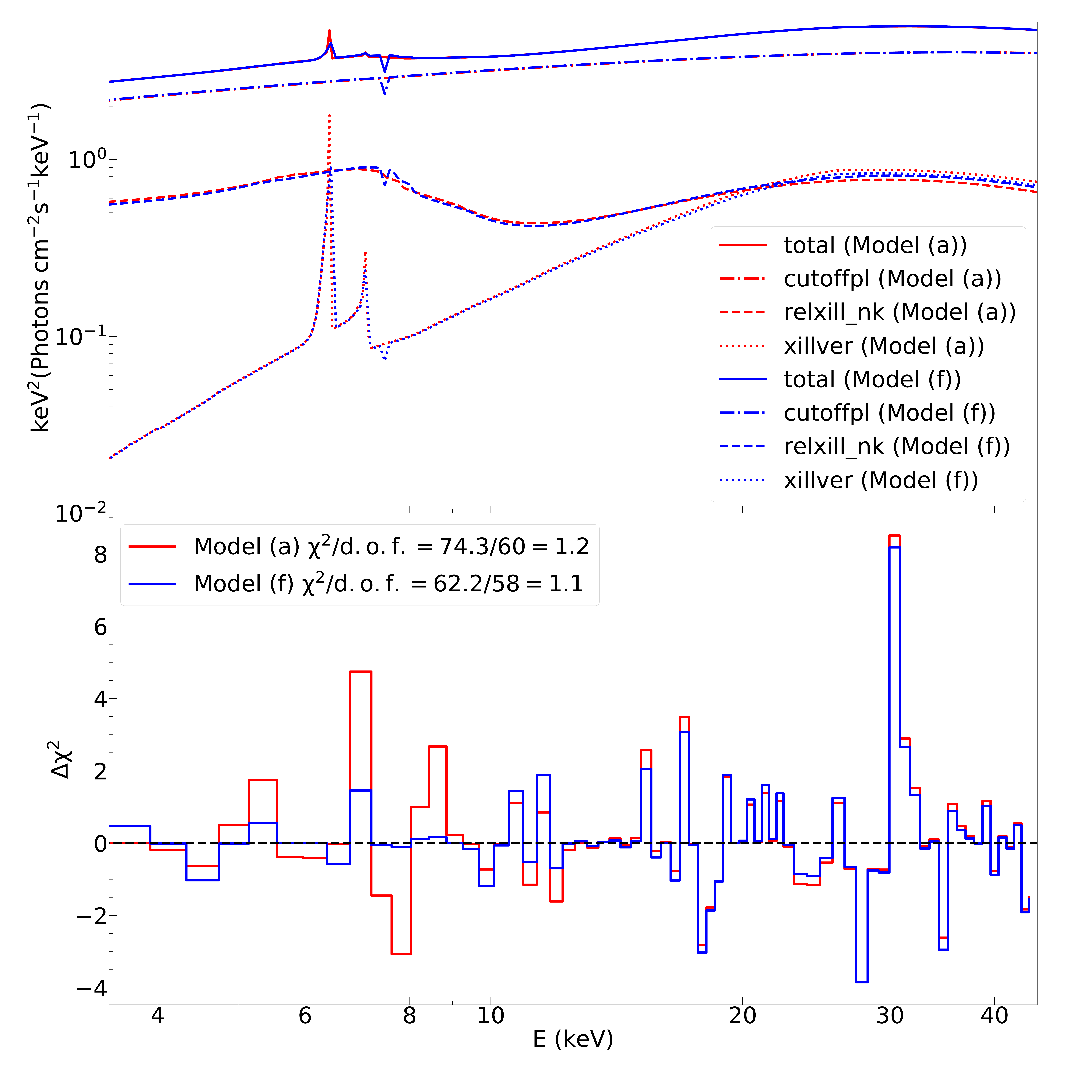}
\end{center}
\vspace{-0.5cm}
\caption{The predicted models and the distributions to $\chi^2$ for model (a) and model (f). While model (a) is a simple model providing a reasonably good $\chi^2$, it cannot fit well the 6-9~keV region. \label{a_vs_f}}
\end{figure}

Thus, the ingredient that makes the strongest effect on estimating the deformation parameter is the choice of model. Specifically, in our case, the choice of model includes whether to include of a Gaussian absorption feature, and whether to use the more physical ``\textsc{Cp}" flavor of incident spectrum, these two are shown to make a difference in the resulting contours; while we do not see a clear evidence of the treatment of Fe abundance in \textsc{xilllver} being an important factor. After comparing to the statistics in \cite{javier_gx339}, we notice that the newer version of \textsc{relxill} only improves the fit moderately, but the freedom to deviate from the Kerr metric solves the discrepancy of Fe abundance in the two reflectors (even if explanations not invoking new physics are definitively preferred~\cite{Jiang:2019xqn}). While models (a) and (b), in which the Kerr solution is not recovered, are simple and provide a reasonably good $\chi^2$, they cannot fit well the 6-9~keV region, especially when compared with models (e) and (f), which eventually turn out to be our preferred models, see Fig.~\ref{a_vs_f}. This shows that \textsc{gabs} is necessary.

As a secondary factor, emissivity profile could also change the results. In our several fits, we only find evidence to support this in Fig.~\ref{fig:contours}(e) and (f), as new region of statistical minimum could possibly show up when the extra d.o.f., $q$, is invited to play. We find no difference with results produced by emissivity profiles (2) and (4) described in Sec.~\ref{fitting}.

Even though model choice could make a huge difference regarding the shape of confidence contours, we still notice that in the fits with decent statistics (except for (c) in Fig.~\ref{fig:contours}), the other 5 contours have a common statistical minimum at very high spin $a_*\sim0.99$ and $\alpha_{13}\sim-1$. The consistent presence of this region of good fit suggests that it may be robust to choice of model selection.

\begin{figure*}
\begin{center}
\includegraphics[type=pdf,ext=.pdf,read=.pdf,width=15cm]{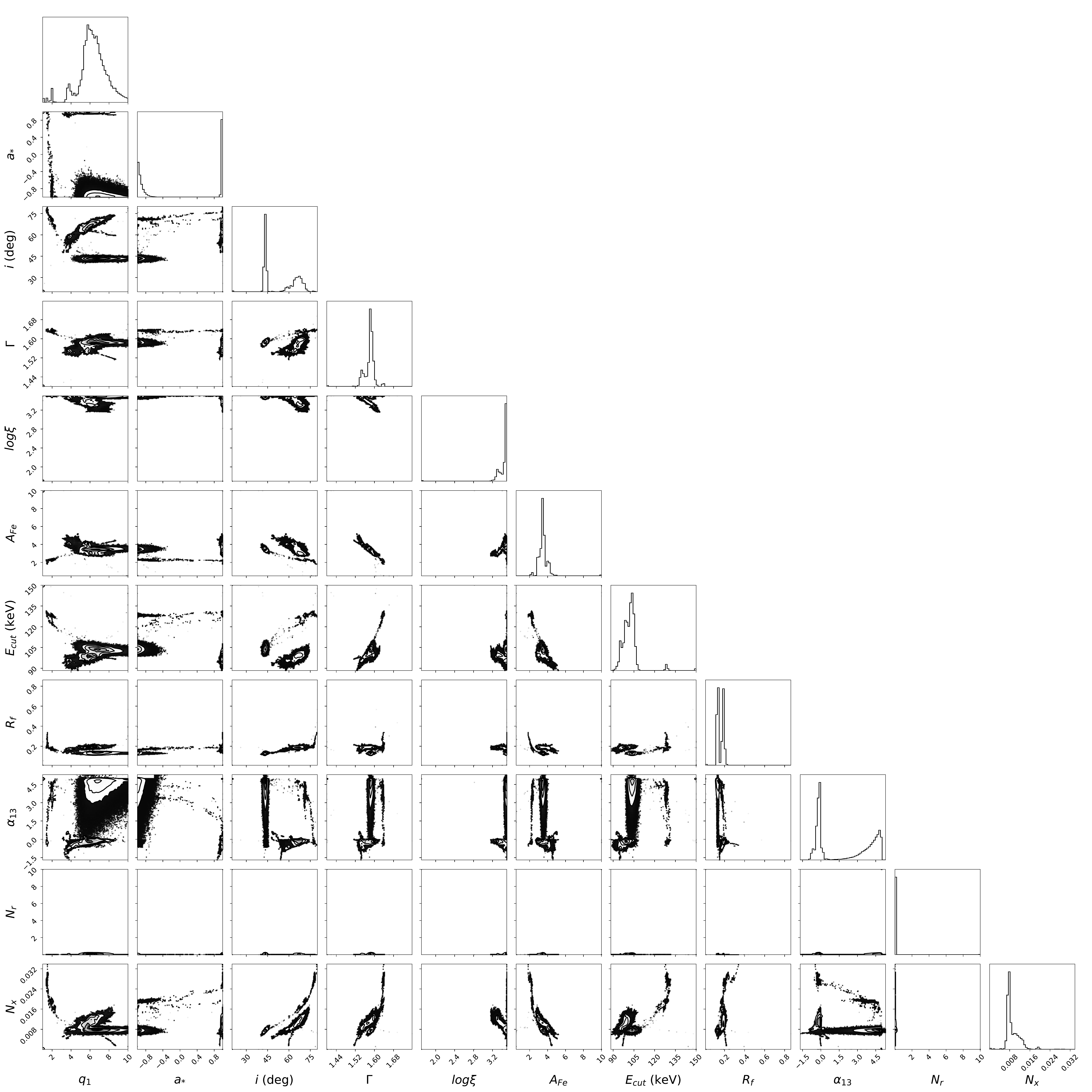}
\end{center}
\vspace{-0.4cm}
\caption{Corner plot for all the free parameter-pairs in Model \textsc{tbabs*(relxill\_nk+xillver)} ($A_{\rm Fe}=1$ in \textsc{xillver}, free $q$) after the MCMC run. This is the model of Fig.~\ref{fig:contours}(a). \label{fig:corner_new_2}}
\end{figure*}

\begin{figure*}
\begin{center}
\includegraphics[type=pdf,ext=.pdf,read=.pdf,width=15cm]{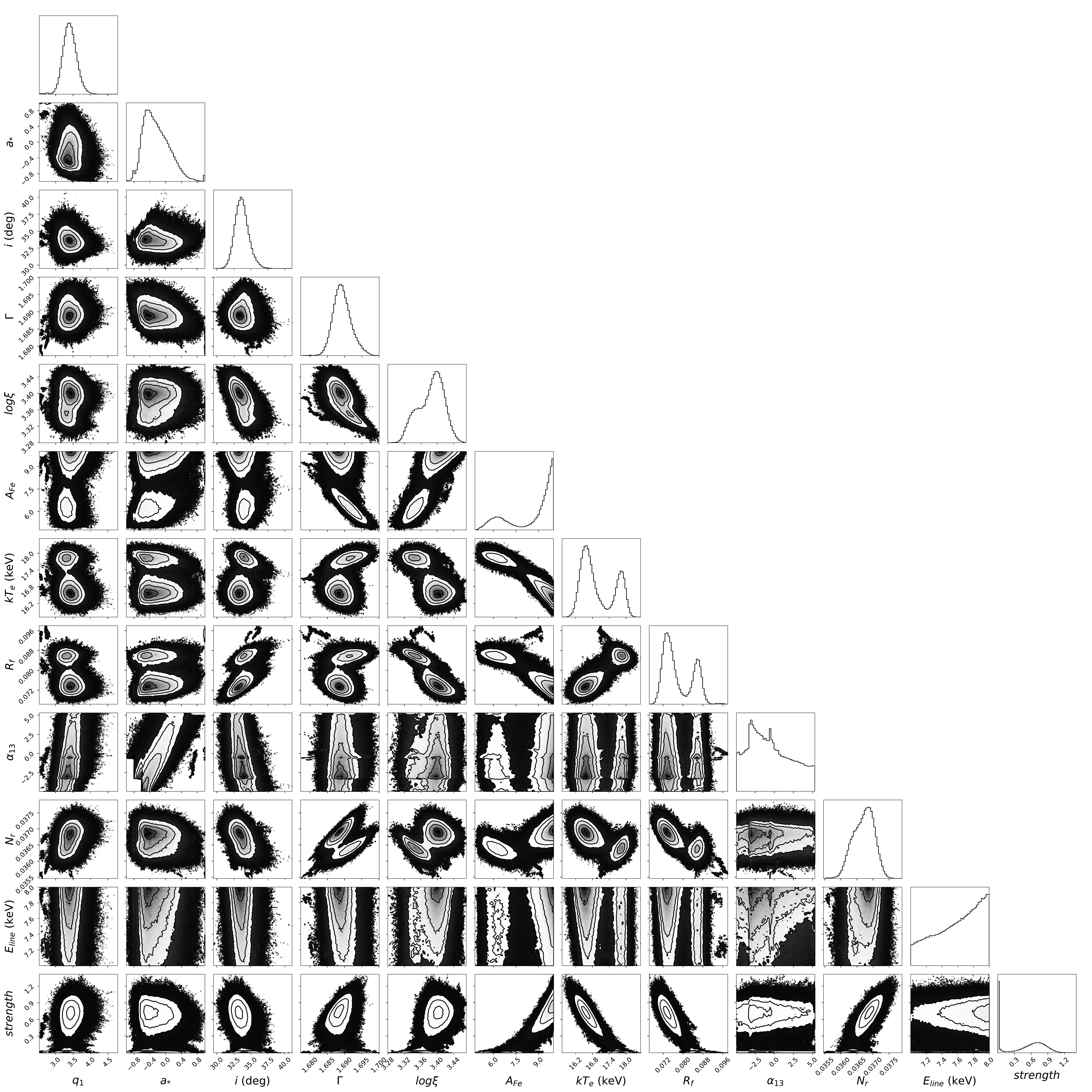}
\end{center}
\vspace{-0.4cm}
\caption{Corner plot for all the free parameter-pairs in Model \textsc{tbabs*relxillCp\_nk} (free $q$) after the MCMC run. This is the model of Fig.~\ref{fig:contours}(d). \label{fig:corner_new_1}}
\end{figure*}

To further investigate the role of the choice of the model, the correlation between $\alpha_{13}$ and the other model parameters, as well as the capability of $\alpha_{13}$ to try to compensate deficiencies in the astrophysical model to describe the observed spectrum, we have performed MCMC analyses of two more models: model~(a) and model~(d) in Tab.~\ref{tab:parameters} and Fig.~\ref{fig:contours}. Both models are simple and provides reasonable good fits. The results from the MCMC analysis are shown, respectively, in Fig.~\ref{fig:corner_new_2} and Fig.~\ref{fig:corner_new_1}. The MCMC analysis can scan better the parameter space than \textsc{steppar} and we can see that, for model~(a), the allowed region is larger and we can recover the Kerr metric. Moreover, the analysis also finds another solution with large deviations from the Kerr geometry and a retrograde disk (which may be rejected on physical grounds as pointed out before).

High spectral resolution data can probably perform better in the selection of the correct astrophysical model and, within a specific model, of a unique solution. While the counting statistics are exquisite with the RXTE dataset, the energy resolution is limited: there are just a few bins in the iron K line region as can be clearly from Figs.~\ref{fig:series_models}, \ref{a_vs_f}, and \ref{fig:weird}. This was the case in the attempt to test the Kerr metric with GRS1915+105. In Ref.~\cite{zhang2019kerr}, we analyzed NuSTAR data, which have a limited energy resolution, and eventually we found impossible to test GR because of the difficulty to select the correct astrophysical model in the presence of possible deviations from the Kerr metric. In Refs.~\cite{zhang19nk}, with the analysis of Suzaku data we were able to select the astrophysical model on the base of the quality of the fits and got quite stringent constraints on possible deviations from the Kerr solution.

There are currently a number of uncertainties in the astrophysical model limiting the ability to perform accurate and precise tests of the Kerr metric using X-ray reflection spectroscopy. The uncertainty in the corona geometry does not permit to have a robust prediction of the emissivity profile for the reflection emission and the commonly adopted power-law or broken power-law emissions are quite crude approximations. The disk is assumed to be infinitesimally thin and the inner edge is set at the ISCO radius, while we know that the disk thickness increases as the mass accretion rate increases and, for high accretion rates, the inner edge of the disk may be at a radius smaller than $R_{\rm ISCO}$. There are still a number of simplifications in the calculations of the reflection emission in \textsc{xillver}: the razor-thinness of the disk, the ionization parameter is constant over radius, the electron density of the disk is throughout fixed and too low for accretion disks of bright black hole binaries, thermal photons from the accretion disk are ignored in the radiative transfer calculations of the reflection emission, etc. Work to remove these simplifications is under way and strictly compulsory if we want to try to use this technique to perform precise tests of GR.

In this paper, we have applied a non-Kerr full X-ray relativistic reflection spectral model to test the Kerr nature of an astrophysical black hole. Since we now have more sophisticated reflection models, like the \textsc{relxill} family, we have taken a step forward and begun analyzing available X-ray data on specific sources available in archive to demonstrate that tests of the spacetime metric around black holes are possible. 

In the present work, we have analyzed a composite spectrum from the detector RXTE PCU-2 on GX~339$-$4 with the highest observed luminosity for hard states. We have only explored the Johannsen deformation parameter $\alpha_{13}$, which has the largest impact on the ISCO radius, and thus on the shape of reflection spectrum, among the deformation parameters of the Johannsen metric. In MCMC analysis with \textsc{tbabs*(relxill\_nk+xillver)*gabs} ($A_{\rm Fe}=1$ in \textsc{xillver}, frozen $q=3$), which eventually turns out to be our preferred model for the considerations discussed in this section, the measurement on the black hole spin parameter $a_*$ and the Johannsen deformation parameter $\alpha_{13}$ are (at 90\% confidence level)
\begin{align}
a_* = 0.92^{+0.07}_{-0.12} \, , \quad 
\alpha_{13} = -0.76^{+0.78}_{-0.60} \, .
\end{align}

The field is in a nascent state, so this work to measure the non-Kerr deformation parameters is still exploratory and limited by the modeling process and fundamentally by our knowledge of black hole astrophysics. The avenues for improvement include better broadband observations with new telescopes coming up in next decade, better modeling of the accretion flow, coronal properties and other aspects of the black hole environment to account for systematic errors, more top-down metrics for physically relevant deformation parameters, and better bottom-up metrics for more comprehensive range of deformation parameters, among others.


\begin{acknowledgments}
J.W. thanks the Cahill Center for Astronomy and Astrophysics at Caltech for support and hospitality during her visit. This work was supported by the Innovation Program of the Shanghai Municipal Education Commission, Grant No.~2019-01-07-00-07-E00035, National Natural Science Foundation of China (NSFC), Grant No.~11973019, and Fudan University, Grant No.~IDH1512060. A.B.A. also acknowledges the support from the Shanghai Government Scholarship (SGS). C.B. and J.A.G. also acknowledge support from the Alexander von Humboldt Foundation. S.N. acknowledges support from the Excellence Initiative at Eberhard-Karls Universit\"at T\"ubingen. J.F.S. has been supported by NASA Einstein Fellowship Grant PF5-160144.
\end{acknowledgments}


\end{document}